\documentclass[12pt,letterpaper]{article}
\pdfoutput=1
\usepackage{amssymb}
\usepackage{amsmath}
\usepackage{amsfonts}
\usepackage{epsfig}
\usepackage{color}
\usepackage{hyperref}
\usepackage[normalem]{ulem}

\usepackage{bm}
 \font\small=cmr10 scaled \magstep0
  
\outer\def\beginsection#1\par{\medbreak\bigskip
      \message{#1}\leftline{\bf#1}\nobreak\medskip
\vskip-\parskip
      \noindent}

\newcommand{\ea}[1]{
\begin{align}
#1
\end{align}
}

\newcommand{\eas}[1]{
\begin{align}
\begin{split}
#1
\end{split}
\end{align}
}

\newcommand{\sea}[1]{
\begin{subequations}
\begin{align}
#1
\end{align}
\end{subequations}
}

\newcommand{\seal}[2]{
\begin{subequations}
\label{#1}
\begin{align}
#2
\end{align}
\end{subequations}
}

\newcommand{\nn}{\nonumber}

\def\Ord{{\cal O}}
\def\D{{\cal D}}
\def\K{{\cal K}}
\def\J{{\cal J}}
\def\P{{\cal P}}

\def\Eq{Eq.~\eqref}

\def\txi{\tilde{\xi}}
\def\tphi{\tilde{\phi}}

\usepackage[titles]{tocloft}

\usepackage{titlesec}
\titleformat*{\section}{\large  \bfseries }
\titleformat*{\subsection}{\normalsize  \bfseries }

\numberwithin{equation}{section}
\begin{document}

%%%%%%%%%%%%%%%%%%%%%%%
\begin{titlepage}
\hfill \hbox{NORDITA-2017-032}
\vskip 1.5cm
\vskip 1.0cm
\begin{center}
{\Large \bf Double-soft behavior of the dilaton of spontaneously broken conformal
invariance 
}
 
\vskip 1.0cm {\large Paolo
Di Vecchia$^{a,b}$,
Raffaele Marotta$^{c}$, Matin Mojaza$^{d}$
} \\[0.7cm] 
{\it $^a$ The Niels Bohr Institute, University of Copenhagen, Blegdamsvej 17, \\
DK-2100 Copenhagen \O , Denmark}\\
{\it $^b$ Nordita, KTH Royal Institute of Technology and Stockholm 
University, \\Roslagstullsbacken 23, SE-10691 Stockholm, Sweden}\\[2mm]
{\it $^c$  Istituto Nazionale di Fisica Nucleare, Sezione di Napoli, Complesso \\
 Universitario di Monte S. Angelo ed. 6, via Cintia, 80126, Napoli, Italy}\\[2mm]
 {\it $^d$  Max-Planck-Institut f\"ur Gravitationsphysik, \\
Albert-Einstein-Institut, Am M\"uhlenberg 1, 14476 Potsdam, Germany}
\end{center}
\begin{abstract}
The Ward identities involving  the currents associated to the spontaneously broken
scale and special conformal transformations are derived and used to determine, 
through linear  order in the two soft-dilaton momenta, the double-soft behavior
 of
scattering amplitudes involving two soft dilatons and any number of other 
particles. It turns out
that the double-soft behavior is equivalent to performing two single-soft limits one after the other.
We confirm the new double-soft theorem
perturbatively at tree-level in a $D$-dimensional conformal field theory model, 
as well as nonperturbatively by using the ``gravity dual'' of ${\cal{N}}=4$ super 
Yang-Mills on the Coulomb branch; i.e. the Dirac-Born-Infeld action 
on AdS${}_5 \times S^5$.
\end{abstract}
\end{titlepage}

%%%%%%%%%%%%%%%%%%%
\tableofcontents
\section{Introduction}
\label{intro}

There are generically two main physical and observable consequences of theories with 
spontaneously broken continuous symmetries; namely i) the appearance of \,
Nambu-Goldstone (NG)
bosons and their dynamics, and ii) the existence of so-called \emph{soft theorems},
 which fix the behavior of scattering amplitudes when the momentum of one or more NG bosons
goes to zero.
They are direct consequences of the Ward identities of the theory.

There are, nevertheless, various important differences between spontaneously breaking an internal
 or a space-time symmetry. {In the case of an internal symmetry, the number of NG
bosons is equal to the number of broken generators, while in the case of a spontaneously
broken space-time symmetry, the number of NG bosons is less~\cite{Nielsen:1975hm}; for instance when conformal symmetry is spontaneously broken to Poincar\'e symmetry
only one NG boson appears, although five generators corresponding to  dilatations and special conformal transformations are broken~\cite{Higashijima:1994zg,Low:2001bw}.

{The two kinds of NG bosons also differ in their soft behavior:
In the case of a spontanously broken internal symmetry, amplitudes involving 
the NG bosons vanish when 
the momentum of one of the NG bosons goes to zero.
A famous example is the non-linear $\sigma$-model (NLSM) describing 
the low-energy behaviour of $SU(n) \times  SU(n)$ theory spontaneously 
broken to the vectorial subgroup $SU(n)$.}
These zeroes are in the literature called Adler zeroes and their discovery, 
purely based on current algebra, dates back to the 1960s~\cite{Adler:1964um, Adler:1965ga,Weinberg:1966gjf}.

The situation is different for the NG boson of spontaneously broken 
conformal invariance, {called \emph{the dilaton}}~\footnote{
It is difficult to give a historical account of this case,
as the early literature from the 1960s on the subject
goes in many directions, not immediately relevant to us.
Let us mention, however, that to our knowledge G. Mack is the first 
that explicitly discusses the dilaton soft behavior and provides its 
leading soft theorem in Ref.~\cite{Mack:1968zz}, while its subleading 
behavior is implicit in the work by D. Gross and J. Wess in Ref.~\cite{Gross:1970tb}.
In these papers earlier references on the realization of conformal 
symmetry in nature is also given, among which the works of 
F. G\"ursey~\cite{Gursey:1956zzb}, J. Wess~\cite{Wess60}, and H. Kastrup~\cite{Kastrup:1962zza}
were frequently cited as well as the early review by T. Fulton, 
F. Rohrlich and L. Witten~\cite{Fulton:1962bu}.
The seminal papers by Callan~\cite{Callan:1970yg}, Coleman 
and Jackiw~\cite{Coleman:1970je} diminished these works to some extend, as 
it was realized that conformal invariance is anomalous in the quantum theory, 
especially that of strong interactions.
The dilaton has reappeared in a more modern context 
in phenomenological models for electroweak symmetry breaking, 
inflationary cosmology, as well as in condensed matter applications.
The modern take on the dilaton soft theorems, especially in the context of 
the recent S-matrix program, were put forward recently in 
Refs.~\cite{Boels:2015pta,Huang:2015sla, DiVecchia:2015jaq}.
}.
In this case the amplitude involving a number of  dilatons together with other
 particles does not vanish when the momentum, $q$, of one dilaton goes 
to zero,
but is fixed in terms of the amplitude without the soft dilaton;
i.e. the dilaton has a nonvanishing soft theorem.
Specifically, since the dilaton couples linearly to the trace of the
energy-momentum tensor, it couples in particular linearly to the mass 
of any massive particles. Therefore there is a nonzero, and in fact divergent, 
{universal} 
contribution to the amplitude associated to the emission of a zero-momentum
dilaton from any massive particle, in complete analogy to the emission of 
soft photons~\cite{Low:1958sn}  and gravitons~\cite{Weinberg:1964ew}.
But moreover, it turns out that the regular part of the
dilaton soft behavior at orders $q^0$ and $q^1$, which is not associated to 
emission from external legs,  is nonzero and 
also fixed universally. This follows from  the
Ward identities of spontaneously broken conformal
 invariance~\cite{Mack:19z,Boels:2015pta,Huang:2015sla,DiVecchia:2015jaq}, 
and applies to both massive and massless particle interactions, as emphasized in Ref.~\cite{DiVecchia:2015jaq}.
This of course applies to conformal theories that are not 
anomalous~\footnote{In generic quantum field theory models of dilatons, the 
presence of the trace anomaly introduces a mass for the dilaton, which only in 
certain  cases can be controllably small~\cite{Antipin:2011aa}.},
and this was in particular tested in the impressive work in Ref.~\cite{Bianchi:2016viy} in the Coulomb branch
of $\mathcal{N}=4$ super Yang-Mills, both perturbatively through one-loop 
and non-perturbatively by considering the one-instanton effective action.
In the same work constraints on dilaton effective actions, new 
non-renormalization theorems, as well as a recursive proof of conformal 
invariance following scale invariance of amplitudes, were all given utilizing 
the soft dilaton theorem of Ref.~\cite{DiVecchia:2015jaq}.

In the case of NLSM-type theories
also the double-soft behavior has
been studied~\cite{Weinberg:1966gjf,Dashen:1969ez,ArkaniHamed:2008gz,Kampf:2013vha,Low:2015ogb, Du:2015esa}.
 In particular, it has been shown that  the amplitude for the emission of 
any even number of NG bosons does no longer vanish when the momenta of two of them go simultaneously to zero. Instead, it is fixed
in terms of the amplitude without the two NG bosons with vanishing momenta and of the structure constants of the group in consideration.

In this work we detail the main physical consequences of a spontaneously broken conformal 
theory.
While property i) has been studied intensively in the literature, little attention has been given to 
property ii), and this is the main purpose of this work. Our main new result is the derivation of
 the double-soft theorem for the NG boson of spontaneously broken conformal symmetry, i.e. 
\emph{the dilaton}.~\footnote{
The subject of single-, double-, and multi-soft theorems has received
much interest in recent years due to their proposed relations with asymptotic symmetries put forward recently by A. Strominger
~\cite{Strominger:2013lka,Strominger:2013jfa}, 
and many papers following.
(In particular, the relation between the soft dilaton and asymptotic symmetries was recently discussed in Ref.~\cite{Campiglia:2017dpg}.)
This has additionally lead to the discovery of many new soft theorems in
 both field and string theory, and has lead to new developments in the 
context of the S-matrix and effective field theory programs,
where of particular relevance we should point out Ref.~\cite{Luo:2015tat, Bianchi:2016viy}.
A comprehensive list of references to this literature can 
be found in Strominger's recent lecture notes in Ref.~\cite{Strominger:2017zoo}.
}

We prove that the soft theorem factorizes any amplitude involving two soft 
dilatons through subleading order in the  two soft 
momenta. We see that, also
 in this case,  the 
double-soft behavior of the dilaton differs from that of the NG bosons  of a 
spontaneously broken internal symmetry. It turns out that
the double-soft behavior of the dilaton, obtained from the Ward identities for the scale and special conformal
transformations, is equivalent to the one obtained by making two single-soft limits one after the other.
This particular form of the double-soft theorem allows us additionally to conjecture an any-multiplicity soft dilaton theorem.

The paper is organized as follows. In Sections  \ref{CFT} and \ref{hidden} we summarize some
general properties of conformal field theories in $D$-dimensional space-time.
In Sect .
\ref{singlesoft} we discuss the Ward identities that follow from the 
conservation of the 
N{\"{o}}ther currents associated with  the scale and special conformal 
transformations. Then,
in a first subsection, we  derive their implication for the scattering amplitude involving
a single current  and an arbitrary number of other states, while, in the two 
subsequent 
subsections, we specify our analysis first to the current corresponding to a scale 
transformation and then to that corresponding to a special
conformal transformation. Sect. \ref{doublesoftgeneral}  is devoted to the case of 
the Ward identities for amplitudes involving two N{\"{o}}ther currents. 
Then, in the first subsection,
we discuss the case of two dilatation currents, and in the second subsection, the case 
of one dilatation current and one current associated to a special conformal 
transformation. In Sect. \ref{multisoft} we show that the double-soft behavior, derived in
Sect. \ref{doublesoftgeneral}) from the Ward identities, can be equivalently obtained 
by performing two
consecutive soft limits, one after the other, and we conjecture that the same behavior  
is also valid in the case of a multi-soft limit. In Sect. \ref{dilatonamplitude} we check 
the previously derived  double-soft behavior with specific amplitudes of a $D$-dimensional
conformal scalar theory that has been recently studied in the literature and of the 
``gravity dual'' of ${\cal{N}}=4$ super Yang-Mills on the Coulomb branch.
{Finally, in the Appendix we give some detail on the calculation of the soft 
behavior in the ``gravity dual'' of ${\cal{N}}=4$ super Yang-Mills 
on the Coulomb branch.}

We would like to add a note of caution: 
The dilaton discussed in this paper should not be confused
with the `gravity dilaton' appearing in the literature on theories of 
(super)gravity and string theory, where it is parametrizing the spin zero 
 mode of the gravitational field. This gravity dilaton also has a soft theorem
similar to the NG dilaton discussed in this work, which was first studied in 
Ref.~\cite{Ademollo:1975pf, Shapiro:1975cz},
where its leading behavior was determined, and recently its subleading behavior 
was also shown to be 
fixed~\cite{DiVecchia:2015jaq, DiVecchia:2015oba, DiVecchia:2016amo,
 DiVecchia:2016szw}.
But the two soft theorems are not equal, although very 
similar~\cite{DiVecchia:2015jaq}, and there is still a lack of rigorous 
understanding of the relation between the two.

\section{Prelude}
\label{CFT}
To set our notations, we start by briefly reviewing aspects of conformal symmetry and its representations in field theory. 
For more details, we refer to the seminal works in 
Ref.~\cite{Mack:1969rr,Callan:1970ze} as well as the textbook in Ref.~\cite{DiFrancesco:1997nk}.

The conformal group is the group that leaves the metric invariant up to a scale $g_{\mu \nu}(x) \to \Lambda(x) g_{\mu \nu}(x)$ and can be considered an extension by dilatations and special conformal transformations of the Poincar\' e group, which belong to $\Lambda(x) =1$. The group is locally isomorphic to $SO(D,2)$, where by $D$ we denote the number of space-time dimensions. 
Infinitesimally, the group transforms space-time coordinates and fields as follows
\ea{
\begin{split}
x^\mu \to {x'}^\mu &= x^\mu + \epsilon^{MN} f_{MN}^\mu (x) \\[2mm]
\Phi(x) \to {\Phi'}(x) &= \Phi(x) + i \epsilon^{MN} \Gamma_{MN}(x) \Phi(x) 
\end{split}
\label{infiniconformaltransformations}
}
where $\epsilon_{MN}$ are infinitesimal parameters and $f_{MN}^\mu$ are functions obeying
\ea{
\partial^\mu f_{MN}^\nu + \partial^\nu f_{MN}^\mu = \frac{2}{D} g^{\mu \nu} \partial_\rho f_{MN}^\rho \, .
\label{fMN}
}
$\Gamma^{MN}$ are the $(D+1)(D+2)/2$ conformal generators, so that $\Gamma^{MN}$ is imaginary and antisymmetric in $M, N = 0, \ldots , D+1$. We consider in this work the flat space limit and take $g^{\mu \nu} \to \eta^{\mu \nu} = {\rm diag}(-1,+1, \ldots)$, and $D> 2$.

It is useful to decompose the conformal transformations and generators into translations, Lorentz transformations, dilatations and special conformal transformations.
First consider the solutions for $f_{MN}$:
 \ea{
 \begin{split}
  f_{D, \mu}^\rho (x) &= \eta_{\mu}^\rho \, , \qquad
 f_{\mu \nu}^\rho(x) = \eta^\rho_\mu x_\nu - \eta^\rho_\nu x_\mu \, ,
 \\
  f_{D+1, D}^\rho (x) &= x^\rho \, , \hspace{2mm}
 f_{D+1, \mu}^\rho (x) = 2 x_\mu x^\rho - \eta_\mu^\rho x^2 
 \end{split}
 \label{fMNx}
 }
where, $\mu, \nu, \rho = 0, \ldots, D-1$ are the space-time indices. The corresponding generators read:
\ea{
\Gamma^{D, \mu} &= \mathcal{P}^\mu 
= i \partial^\mu
\, , 
\hspace{25mm}
\Gamma^{\mu \nu} = \mathcal{J}^{\mu \nu} 
= - i (x^\mu \partial^\nu - x^\nu \partial^\mu ) - \mathcal{S}^{\mu \nu}
\, , 
\label{generators}
\\
\Gamma^{D+1, D} &= \mathcal{D} 
= i(d_\Phi + x_\mu \partial^\mu)
\, , \quad
\Gamma^{D+1, \mu} = \mathcal{K}^\mu 
=i(2 x^\mu x_\nu \partial^\nu - x^2\partial^\mu + 2d_\Phi x^\mu) + 2x_\nu \mathcal{S}^{\mu \nu}  
\,  ,
\nonumber
}
where $\mathcal{P}^\mu$ are the generators of translation, $\mathcal{J}^{\mu \nu}$ are the generators of Lorentz transformations with $\mathcal{S}^{\mu \nu}$ corresponding to the spin angular momentum operator, $\mathcal{D}$ is the generator of dilatation, and $\mathcal{K}^\mu$ are the generators of special conformal transformation. The coefficient $d_\Phi$ denotes the scaling dimension of the field $\Phi$.
The generators obey the commutation relations:
\eas{
[ \J^{\mu \nu}, \J^{\rho \sigma} ] &= i (\eta^{\mu \rho}\J^{\nu \sigma} + \eta^{\nu \sigma}\J^{\mu \rho} 
-
\eta^{\mu \sigma}\J^{\nu \rho}
-
\eta^{\nu \rho}\J^{\mu \sigma}
)  \\
[\P^\rho, \J^{\mu \nu} ] &= i (\eta^{\rho \nu} \P^\mu - \eta^{\rho \mu} \P^\nu )
 \\
[\K^\rho, \J^{\mu \nu} ] &= i (\eta^{\rho \nu} \K^\mu - \eta^{\rho \mu} \K^\nu )
 \\
[\K^\mu, \P^{\nu} ] &= 2 i (\J^{\mu \nu} - \eta^{\mu \nu} \D )
 \\
[\D, \P^{\mu} ] &=  - i \P^\mu 
 \\
[\D, \K^{\mu} ] &=   i \K^\mu  \, ,
}
with all other commutators vanishing.

The currents associated to the conformal generators can be constructed by
varying the conformal invariant action as in 
\Eq{infiniconformaltransformations} assuming that the
infinitesimal parameters $\epsilon_{MN}$ are arbitrary functions of $x$. In
this way, for the conformal group, one can get:
\ea{
\delta S =  
\int d^D x \, \epsilon^{MN} (x) (\partial_\mu J^\mu_{MN})
=
 \int d^D x \, \epsilon^{MN} (x) \partial^\mu \left(
f^\nu_{MN} T_{\mu \nu}  \right)
\label{dMNS}
}
where $T_{\mu \nu}$ is the improved energy-momentum tensor.
Using \Eq{fMN}, it turns out that the N{\"{o}}ther current is conserved if the
improved energy-momentum tensor is conserved and traceless:
\ea{
\partial^\nu T_{\mu \nu} = T_\mu^\mu =0 
\label{contrace}
}
 when the classical equations of motion are satisfied.
It is easy
to see by \Eq{fMNx} that the currents 
$J_{\mu \nu}^\rho$ and $J_{D,\mu}^\rho$ are conserved independently 
of the zero-trace condition $T_\mu^\mu=0$, 
since $\partial_\rho f_{\mu \nu}^\rho = \partial_\rho f_{D, \mu}^\rho = 0$.
The currents $J_{D+1,\mu}^\rho$ and $J_{D+1,D}^\rho$, on the other hand, 
are only conserved if $T_\mu^\mu=0$.
Specifically,
\sea{
J_{D+1, D}^\mu &= J_{\D}^\mu = x_\nu T^{\mu \nu} \, , 
\hspace{5mm}
J_{D+1, \rho}^\mu = J_{\K,\rho}^\mu = (2 x_\nu x_\rho - \eta_{\rho\nu} x^2 ) T^{\mu \nu}
\label{currents} \\[3mm]
\partial_\mu J_{\D}^\mu &=T^\mu_\mu 
\, , \hspace{21,5mm}
\partial_\mu J_{\K, \rho}^\mu = 2 x_\rho T^\mu_\mu
\label{currentsdivergence}
}
where we 
stress once more
that $T^{\mu \nu}$ is the improved energy-momentum 
tensor.

\section{Hidden conformal symmetry}
\label{hidden}
We consider the situation where conformal symmetry of some underlying conformal field theory is spontaneously broken due to a Lorentz scalar primary operator getting a nonzero vacuum expectation value (vev), i.e.
\ea{
\langle 0 | \mathcal{O}_{\rm scalar} | 0 \rangle = v^{d_{\mathcal{O}}} \, ,
}
where $d_\mathcal{O}$ is the scaling dimension of $\mathcal{O}$ so that $v$ has mass dimension one. The vev $v$ is the only mass scale of the broken theory.
The vacuum remains invariant under Lorentz transformations and translations, 
but dilatations and special conformal transformations are then no longer 
symmetries of the vacuum.

It follows from Goldstone's theorem \cite{Higashijima:1994zg,Low:2001bw} 
that a massless scalar 
state of conformal dimension one (for $D=4$)
appears in the spectrum 
of the broken theory, parametrizing the massless excitations of the vacuum  
generated by the broken symmetry currents. This Nambu-Goldstone (NG) 
boson of spontaneously broken scale invariance is also known as 
\emph{the dilaton}.\footnote{Although $D+1$ generators are broken, only 
one NG boson appears. This mismatch of degrees of freedom is a 
consequence of space-time symmetries being broken, as opposed to 
when global continuous internal symmetries are broken~\cite{Low:2001bw}.}

As a consequence, the dilaton couples linearly to the energy-momentum tensor
\ea{
       T_{\mu \nu} &= - 
       f_\xi
     \left (  \frac{\eta_{\mu \nu} \partial^2  - \partial_\mu \partial_\nu }{D-1}\right )
     \xi(x) + \cdots  \, ,
     \label{Tmunu}
}
where $\xi(x)$ parametrizes the dilaton field and $f_\xi$ is a dimensionful 
constant, thus related to $v$, which can be thought of as the dilaton decay 
constant. The ellipsis $\cdots$ denote term of higher order in the fields, i.e. 
the 
dilaton is the only field that couples linearly to the energy-momentum tensor.

Taking the trace of the above expression and imposing the equation of motions leads to the expression
\ea{
       T_\mu^\mu = f_\xi (- \partial^2 \xi) \ , 
       \label{Tmumu}
}
which is exact on the equations of motion. As expected, the trace of the energy-momentum no longer vanishes in the broken phase.
Instead it is simply parametrized by the equation of motion of the dilaton field. 
The statement can also be reversed; the dilaton equation of motion is described 
by the trace of the energy-momentum tensor.

To better appreciate the latter statement, and to also comment on the occurrence of the trace anomaly in generic quantum field theories, let us consider a generic renormalized action in $D$-dimensions.
It can be described  in a basis of eigenoperators $\Psi_i$ of (renormalized) scaling dimension $d_i$ as follows:
 \ea{
S_0(\mu) = \int d^D x \sum_i  g_i (\mu ) \Psi_i (x) \, ,
}
where $g_i(\mu)$ are renormalized coupling constants at a renormalization scale $\mu$.

The change of the action under dilatations yields by definition the trace of the 
energy-momentum tensor, as can be verified from Eq.~\eqref{dMNS}
using Eq.~\eqref{fMNx}.
Specifically, taking ${x'}_\mu = e^\lambda x_\mu \approx x_\mu +  
\lambda \,  x_\mu$, yields for any action
\ea{
\delta S  = \lambda \int d^D x \, T_\mu ^\mu (x) \, .
}
The explicit variation of the action $S_0$ is, on the other hand, readily derived 
by making a scale transformation of the scale $\mu' = \mu e^{-\lambda}$, as 
well as of the (scalar) operators \mbox{$\Psi_i'(x) = e^{-\lambda d_i } \Psi_i({\rm e}^{-\lambda}x)$}.
Then for infinitesimal transformations, we find at linear order in $\lambda$:
 \ea{
 \delta S_0 =- \lambda \int d^D x \, 
 \sum_i \left ( (d_i -D) g_i(\mu)  + \mu \frac{\partial g_i}{\partial \mu}  \right )
\Psi_i(x) \equiv \lambda \int d^D x \, {T_0}_\mu ^\mu (x) \, ,
\label{T0mumu}
}
where in the second equality we identified ${T_0}^\mu_\mu$.
For marginal operators where the scaling dimension $d_i = D$, the 
first term vanishes, but the $\beta$-functions, \mbox{$\beta_i(g) =\mu \partial g_i/\partial \mu$}, for 
the corresponding coupling constants still contribute to the trace. 
This is the consequence of the trace-anomaly for general quantum field 
theories. In a theory with only marginal operators and where the couplings 
remain unrenormalized, the trace anomaly vanishes. 
An example is $\mathcal{N}=4$  super Yang-Mills theory.

Let us now connect this to our previous expressions for a spontaneously 
broken conformal theory.
It is possible to render the action scale invariant by introducing a conformal 
compensator field~\cite{Goldberger:2008zz}, $\bar{\xi}$, with canonical kinetic 
term and free-field scaling dimension \mbox{$d = (D-2)/2$}
by the following formal replacement:
\ea{
g_i(\mu) \Psi_i(x) \to g_i \left(  \frac{\mu\, v}{{\bar{\xi}}^{\frac{1}{d}}(x)} 
\right ) \left (\frac{{\bar{\xi}}(x)}{v^d} \right )^{\frac{D-d_i}{d}} \Psi_i (x)
}
yielding the Lagrangian
\ea{
\mathcal{L}_0(\mu) \to \mathcal{L}(\mu ) =  
- \frac{1}{2} 
 \partial_\nu \bar{\xi} \partial^\nu \bar{\xi}  
+\sum_i g_i \left(  \frac{\mu\, v}{{\bar{\xi}}^{\frac{1}{d}}(x)} 
\right ) \left (\frac{\bar{\xi}(x)}{v^d} \right )^{\frac{D-d_i}{d}} \Psi_i(x)
\label{Lagra}
} 
It is easy to check that under the transformations
\ea{
~~ {\bar{\xi}} (x) \rightarrow {\rm e}^{-\lambda d} {\bar{\xi}}({\rm e}^{-\lambda} x)~~;
~~ \Psi_i (x) \rightarrow  {\rm e}^{-\lambda d_i} \Psi_i ({\rm e}^{-\lambda} x)~~;~~
\mu \rightarrow {\rm e}^{-\lambda} \mu \,\,,
\label{scaletra}
}
the action, corresponding to the Lagrangian in Eq. (\ref{Lagra}), is left invariant. 
The introduction of the field dependent coupling constants is a formal trick 
and should be understood as an expansion
in the shifted 
(dilaton) field \mbox{$\bar{\xi} = v^d + \xi$}, which is well-defined only in 
the broken phase, i.e.
\ea{
g_i \left (\frac{\mu \, v}{  \bar{\xi}^{\frac{1}{d}}(x)}  \right )  =
g_i (\mu) - \frac{\xi(x)}{ d v^d} \,\,\mu \frac{\partial g_i}{\partial \mu}  +
 \cdots
}
Alternatively, the formal replacement can also be understood through a 
nonlinear realization, by the replacement of the field $\bar{\xi}(x) =
 v^d e^{\sigma(x)/v^d}$.

Now it can be checked that the renormalized low-energy action of the 
broken phase, 
where $\xi \ll v^d$, is given by
\ea{
S(\mu) = S_0 (\mu ) + \int d^D x \, \left (
- \frac{1}{2}\partial_\mu {\xi} \partial^\mu {\xi} 
+ \frac{1}{d} \frac{\xi}{v^d} {T_0}_\mu^\mu + \cdots
\right )
}
 where ${T_0}_\mu^\mu$ was defined in \Eq{T0mumu} and the ellipses $\cdots$ stand for terms of higher order in $\xi/v_d$.
Finally, we see that the equation of motion of the dilaton $\xi$ is given by:
\ea{
d\,  v^d (- \partial^2 \xi ) = {T_0}_\mu^\mu + \cdots
}
This is equivalent to the general expression in Eq.~\eqref{Tmumu}, with the 
identification of the decay constant $f_\xi = d v^d$. We furthermore learn 
that this expression contains the effects of renormalization, or, in other words, 
of the trace anomaly of the theory without the dilaton.
The low-energy effective action of the dilaton can also be obtained by 
integrating out the massive fields in the broken phase, and can be constructed 
using anomaly matching considerations, put  forward
in Ref.~\cite{Schwimmer:2010za}, and since studied in the recent a-theorem 
literature~\cite{Komargodski:2011vj,Elvang:2012st,Elvang:2012yc}.

The simplest example of the above construction is to consider a free massive 
scalar field in $D$ dimensions. Its Lagrangian reads:
\ea{
\mathcal{L}_0 = -  \frac{1}{2}\partial_\mu {\chi} \partial^\mu {\chi} - 
\frac{m^2}{2} \chi^2 \, 
}
with $[\chi] = d = (D-2)/2$.
Introducing the conformal compensator, and defining a dimensionless 
coupling constant $\lambda$ through the relation
$m^2 = v^2 \lambda^{2/d}$, the resulting theory reads:
\ea{
 \mathcal{L} = - \frac{1}{2}(\partial_\mu \chi)^2 - \frac{1}{2} 
(\partial_\mu {\bar{\xi}})^{2} 
- \frac{v^{2}}{2} \left (\frac{{\lambda} {\bar{\xi}} }{v^d} \right )^{2/d} \chi^2 
\label{simplecft}
}
This is a classically scale invariant theory in $D$ dimensions. As we have 
argued, it stays scale invariant in the renormalized theory by substituting 
$\lambda(\mu) \to  \lambda(\mu v/ \bar{\xi}^{1/d}(x) )$. 
This model has been considered in recent works~\cite{Shaposhnikov:2009nk,Armillis:2013wya,Gretsch:2013ooa,Boels:2015pta}, 
where its validity as a quantum conformal theory has been studied (see also the early related work~\cite{Englert:1976ep}). We will later come back to this model for computing tree-level scattering amplitudes of the theory, and thus only its classical scale invariance is of importance to us.

\section{Current algebra and soft theorems from Ward identities}
\label{singlesoft}
An observable consequence in scattering processes of spontaneously broken symmetries is  the so-called \emph{soft theorems}.  These are identities relating S-matrices with NG bosons to S-matrices without the NG bosons, and they exist as a consequence of the Ward identities of the broken symmetry currents.

In this section we detail the relationship between Ward identities of the
broken conformal currents and their implications on scattering amplitudes,
 leading to soft theorems for the dilaton. The main observation of the previous 
section that we will draw on, is that, by Eq.~\eqref{currentsdivergence} 
and Eq.~\eqref{Tmumu}, the divergence of the broken conformal currents are 
parametrized solely by the equation of motion of the dilaton, i.e.
\ea{
\partial_\mu J_{\D}^\mu &= f_\xi (-\partial^2 \xi)
\, , \qquad
\partial_\mu J_{\K, \rho}^\mu = 2  f_\xi x_\rho (-\partial^2 \xi) \, .
\label{dilatoncurrents}
}
The dilatation current is broken by a dimension $d+2$ operator, while the special conformal transformation currents are broken by dimension $d+1$ operators, where \mbox{$d= [\xi] = (D-2)/2$}. 
In both cases, the dimensions are below the space-time dimension $D=2d +2$, and the currents can thus be considered \emph{partially conserved}~\cite{Treiman:1986ep}. It is due to this that we can proceed and formulate a current algebra for the spontaneously broken symmetries in analogy to the PCAC method.

 The main object one must study to derive low energy theorems is the matrix element
 \ea{
T^\ast \langle 0| J_1^{\mu_1} (y_1) \cdots J_m^{\mu_m} (y_m) \phi_1 (x_1) \cdots
 \phi_n (x_n) |0 \rangle
 \label{matrixelements}
 }
where  $J_i$ represent some broken symmetry currents, $\phi_i$ are
generic fields with scaling dimension $d_i$, which for simplicity will all be taken to be scalars,
and $T^\ast$ denotes the Lorentz covariantized time-ordered product, which for our concern implies that derivatives act outside of the time ordering.
  This is a modified definition of the usual T-product, which importantly  
leads to the removal of the so-called \emph{Schwinger terms} 
in the Ward identities, when the currents are partially 
conserved~\cite{Gross:1970ee,Treiman:1986ep}.

  It will be useful to define the Fourier transformed field operators:
 \eas{
\tilde{ J}_{i}^\mu (q) &= \int d^D x\ e^{-i q\cdot x} J_{i}^\mu (x) \,  ,\quad  \text{for } i = 1, \ldots, m \\
\tilde{ \phi}_i (k_i) &= \int d^D x\ e^{-i k_i\cdot x_i} \phi_i (x_i) \, ,\quad  \text{for } i = 1, \ldots, n
}
It is also useful to remember that the charge associated to a current is given by:
\ea{
Q_{i} = \int d^{D-1} \mathbf{x}\ J_{i}^0 (t,\mathbf{x}) = \int d^D x \ J_{i}^0 (x) \delta(x^0 - t)
}
and that the infinitesimal symmetry transformation of a field associated with a conserved current is given by the equal-time commutator:
\eas{
\delta_{i} \phi (x) =  i [ Q_{i} , \phi (x) ] 
&
= i\int d^D y \  [ J_{i}^0(y), \phi(x) ] \delta(x^0 - y^0)
\label{deltaphi}
}
By Eqs.~\eqref{infiniconformaltransformations} and \eqref{deltaphi}
the charges and the generators are simply related by:
\ea{
[Q_{i} , \phi(x)] = \Gamma_{i} (x) \phi(x) \, .
\label{QMN}
}

A basic assumption for current algebra is that we can make use of the following distributional identification
even for partially conserved currents:
\ea{
 [ J_{i}^0(y), \phi(x) ] \delta(x^0 - y^0) : =[ Q_{i} , \phi (x) ]  \delta^{(D)} (x-y)  
 \label{distributional}
}
This identity assumes that there are no boundary terms that vanish upon integrations, which are the would-be Schwinger terms. We assume that such terms vanish, as is generally true for partially conserved currents when considered in $T^\ast$-ordered correlation functions.

This assumption becomes useful when considering the derivative of the matrix-element with respect to the current coordinates. For instance:
\ea{
\partial_\mu^x 
T^\ast \langle 0| J_1^\mu (x) \Phi (y) |0 \rangle
&= T^\ast \langle 0| \partial \cdot J_1 (x) \Phi (y) |0 \rangle
+ \delta^{(D)}(x-y) \langle 0| [Q_1 ,\Phi (y)] |0 \rangle
}
where the second term on the {right-hand} side {arises} 
from taking the derivative 
of the step-function $\theta$, and we used the distributional identity \Eq{distributional}.

In the following subsections we study the Ward identity implications for the case of spontaneously broken symmetries, and in particular the specific cases of theories with broken dilatation and special conformal transformations.

\subsection{Single-soft Ward identity: General treatment}
\label{singlesoftgeneral}
Considering one 
derivative
acting on the matrix element of the $T^\ast$-ordered product of operators with one current insertion, we get:
\ea{
&\partial_\mu^x
T^\ast \langle 0| J^\mu (x)  \phi (x_1) \cdots
 \phi (x_n) |0 \rangle
 \label{matrixderivative}
 \\
& = 
 T^\ast \langle 0|\partial_\mu J^\mu (x) \phi (x_1) \cdots
 \phi (x_n) |0 \rangle
 - i 
 \sum_{i=1}^{n} \delta^{(D)}(x-x_i) 
  T^\ast \langle 0| \phi (x_1)  \cdots
  \delta \phi (x_i ) \cdots
 \phi (x_n) |0 \rangle
   \nonumber
}
If $J^\mu$ parametrizes an unbroken symmetry, its divergence vanishes, and 
thus the first term on the right-hand side is zero, leading to the usual Ward 
identity of conserved currents.

If the symmetry is, on the other hand, spontaneously broken one may proceed 
in two different ways: One can  either define and work with 
a new current, whose divergence also vanishes 
(as we will briefly explain below)
 or parametrize the divergence of the current in terms of the associated NG boson. In these notes we are taking the latter approach.

Due to Eq.~\eqref{dilatoncurrents}, we assume the divergence of the current to 
be parametrized in terms of the NG boson, $\xi$, i.e. the dilaton, as follows:
\ea{
\partial_\mu J^\mu (x) = g_J (x) (- \partial^2) \xi(x)
\label{gJ}
}
where $g_J$ is some function that is determined by the broken symmetry current, $J$.
The function $g_J$ may at most be of dimension less than $D$ to obey partial 
conservation.
Furthermore, if $g_J$ satisfies $\partial^2 g_J = 0$, which is the case 
for dilatations and special conformal transformations, then, as mentioned 
before, we may define a new 
conserved  current 
$j^\mu (x) = J^\mu + g_J(x) (\partial^\mu \xi) - (\partial^\mu g_J) \xi(x)$, 
and work with this instead, by standard methods.

Now, let us consider Eq.~\eqref{matrixderivative} for small transferred momentum of the current.
We will in general assume that the Fourier transform of the correlation functions in \Eq{matrixelements} have no pole in the momentum variables of the current. This implies that the left-hand side of Eq.~\eqref{matrixderivative} vanishes in the soft limit of transferred momentum, i.e.
\eas{
i q_\mu \langle 0| \tilde{J}^\mu (q)  \phi (x_1) \cdots
 \phi (x_n) |0 \rangle
= 0 + \Ord (q)
\label{assumption}
}
This leads to what we call the \emph{single-soft Ward identity}:
\eas{
&\int  d^D x\,  {\rm e}^{-i q x} 
T^\ast \langle 0|\partial_\mu J^\mu (x) \phi (x_1) \cdots
 \phi (x_n) |0 \rangle
 \\
 &
 = 
 - \sum_{i=1}^{n}   {\rm e}^{-i q x_i} 
   T^\ast \langle 0| \phi (x_1)  \cdots
 \Gamma_{J}(x_i) \phi (x_i ) \cdots
 \phi (x_n) |0 \rangle
 + \Ord(q)
 \label{singlesoftWI}
}
where we used the relation in Eq.~\eqref{QMN} between the charge 
commutator and the infinitesimal generators, which, on more general ground, 
may induce a 
linear combination of other fields, and this should be understood 
implicitly.

The Fourier transform on all variables of \Eq{singlesoftWI} leads to the 
momentum space version of the single-soft Ward identity:
\ea{
\tilde{g}_J (q) \left ( q^2 \langle \txi(q) \tphi(k_1) \cdots \tphi(k_n) \rangle \right )
= - \sum_{i=1}^n \tilde{\Gamma}_{J}(k_i+q) 
\langle  \tphi(k_1) \cdots \tphi(k_i+q)  \cdots \tphi(k_n) \rangle
+
\Ord(q)
\label{singlesoftWI2}
}
where by $\langle \cdots \rangle$ we denote the Fourier transform of the 
$T^\ast$-ordered matrix element, and we remember that the Fourier transform 
of functions 
of $x$ are operators in the dual momentum space.
Amputating the correlation functions reduces the expression further. 
The amputated correlation function is defined as:
\ea{
\langle \tphi_1 \cdots \tphi_n \rangle_{\rm amp} = \frac{\langle \tphi_1 \cdots \tphi_n \rangle}{
\langle \tphi_1 \tphi_1 \rangle
\cdots
\langle \tphi_n \tphi_n \rangle } \, .
}
The two-point correlator (the propagator) of a scalar field reads:
\ea{
\Delta_i (k) \equiv \langle \tphi_i (k) \tphi_i (k) \rangle  = \left ( \frac{(-i)}{k^2 + m^2} \right )^{{\frac{D}{2}-d_i}}
\label{Deltai}
}
where $m$ is the mass of the scalar field, $d_i$ is its scaling dimension, $D$ is the number of space-time dimensions, and we defined $\Delta_i$.

Since $\xi$ is massless and $[\xi] = (D-2)/2$, it follows from \Eq{singlesoftWI2} that
\eas{
&i \prod_{i=1}^n \Delta_i(k_i)\, 
\tilde{g}_J (q)  \langle \txi(q) \tphi(k_1) \cdots \tphi(k_n) \rangle_{\rm amp} 
\\
&
= \prod_{j\neq i}^n \Delta_j(k_j)
\sum_{i=1}^n \tilde{\Gamma}_{J}(k_i+q) 
\Delta_i(k_i + q)
\langle  \tphi(k_1) \cdots \tphi(k_i+q)  \cdots \tphi(k_n) \rangle_{\rm amp} 
+\Ord(q)
}
It is useful to define the commutator of $\tilde{\Gamma}_{J}$ with the propagator as the propagator multiplying a new operator $\tilde{F}_{J}(k_i+q, m_i)$, i.e.:
\ea{
\left [
\tilde{\Gamma}_{J}(k_i+q) \, , \
\Delta_i(k_i+q)
\right ] =
\Delta_i(k_i+q)
\tilde{F}_{J} (k_i + q, m_i )
\label{FJ}
}
This allows us to finally write the soft Ward identity as an identity among amputated correlators:
\ea{
&i
\tilde{g}_J (q)  \langle \txi(q) \tphi(k_1) \cdots \tphi(k_n) \rangle _{\rm amp} 
\label{singlesoftWI-FT}\\
&
= \sum_{i=1}^n
\left (\tilde{F}_{J}(k_i+q, m_i) + \tilde{\Gamma}_{J}(k_i+q)  \right )
\langle  \tphi(k_1) \cdots \tphi(k_i+q)  \cdots \tphi(k_n) \rangle_{\rm amp} 
+\Ord(q)
\nonumber 
}
It shows explicitly the relation between the correlation functions with a 
soft Nambu-Goldstone boson inserted 
and the correlation functions without the Nambu-Goldstone boson. 
The expansions in $q$ of the right-hand side should be considered with care.

We can proceed further and derive the consequences of this identity on amplitudes.
According to LSZ reduction, the amplitude is the on-shell residue of correlation functions in Fourier space, or equivalently in terms of the amputated correlation functions it is the on-shell $T$-matrix element of those functions. Therefore the previous expression yields the relation:
\ea{
&i \tilde{g}_J (q)  
 \delta^{(D)}(\textstyle{\sum_i} k_i + q) T_{n+1} (q; k_1, \ldots, k_n)
\\
&
=
\sum_{i=1}^n
\left (\tilde{F}_{J}(k_i+q, m_i) + \tilde{\Gamma}_{J}(k_i+q)  \right )
\delta^{(D)}(\textstyle{\sum_i} k_i + q) T_{n} (k_1, \ldots, k_i +q, \ldots k_n)
+\Ord(q)
\nonumber
}
To remove the delta functions on both sides, we need to commute the momentum-conserving delta-functions through the $\tilde{F}_{J}$ and $\tilde{\Gamma}_{J}$ operators. We assume that this commutator is a function multiplying the delta-function over the momenta and  we thus define:
\eas{
\left [ \tilde{F}_{J}(k_i+q, m_i)  \, , 
\delta^{(D)}(\textstyle{\sum_i} k_i + q)
\right ] &= f_{J} (k_i +q , m_i) \delta^{(D)}(\textstyle{\sum_i} k_i + q)
 \\
\left [ \tilde{\Gamma}_{J}(k_i+q)  \, , 
\delta^{(D)}(\textstyle{\sum_i} k_i + q)
\right ] &= \gamma_{J} (k_i +q ) \delta^{(D)}(\textstyle{\sum_i} k_i + q)
\label{fg}
}
In the next sections we will see that this assumption is satisfied in the case of the  scale and special conformal transformations.

The soft-identity for amplitudes now reads:
\eas{
i \tilde{g}_J (q)  T_{n+1} (q; k_1, \ldots, k_n)
=
\sum_{i=1}^n
\Big [&f_{J}(k_i + q, m_i) +  \gamma_{J} (k_i +q ) 
+\tilde{F}_{J}(k_i+q, m_i) 
 \\
&+ \tilde{\Gamma}_{J}(k_i+q)  
\Big ]
T_{n} (k_1, \ldots, k_i +q, \ldots k_n)
+\Ord(q)
}
where momentum conservation is implicit on both sides. More 
precisely, the identity holds once momentum conservation is imposed to 
fix the same momentum on both sides of the equation. To make this 
statement explicit in our expression, we introduce the notation for the 
$n$th momentum:
\ea{
\bar{k}_n = - \sum_{i=1}^{n-1} k_i - q
\label{kn}
}
meaning that one hard momentum is kept fixed.
The expansion in $q$ on both sides should be done carefully, once the 
functions and operators are specified. Whether this leads to a soft theorem 
depends on the Fourier transform of $g_J$ which may be an operator valued 
function acting on the dual momentum variables.

\subsection{Soft Ward identity of the dilatation current, $J_\D^\mu$}
\label{singlesoftdilatation}
We consider the construction in the previous section for the specific case
of dilatations $J^\mu = J_\D^\mu$ and $\Gamma_\D = \D$.
Following the notation of Sec.~\ref{CFT} we have
\ea{
\D_\varphi (x) \varphi(x)= i(d_\varphi + x^\mu \partial_\mu ) \varphi(x) \, , 
\quad
 \partial_\mu J_\D^\mu = T_\mu^\mu = f_\xi (-\partial^2) \xi \, , 
\label{SingleDilatation}
}
where $\varphi$ is any field and $d_\varphi$ is its scaling dimension.
This defines $g_\D(x) = f_\xi$.
The Fourier transforms are:
\ea{
\tilde{g}_\D =  f_\xi \, , \quad \tilde{\D}_{i}(k) = i (d_i - D - k \cdot \partial_k )
}
and the commutator with the scalar propagator reads:
\ea{
&\left [
\tilde{\D}_{i}(k_i+q) \, , \,
\Delta_i(k_i + q)
\right ] 
=
\Delta_i(k_i + q)
\left [
i (D-2d_i)\left ( 1- \frac{m_i^2}{(k_i+q)^2 + m_i^2} \right )\right ]
}
This defines the operator $\tilde{F}_i(k_i +q, m_i)$, which is simply a function 
because $\D_i$ is a linear operator. In the massless case it is simply a number 
$\tilde{F}_i(k_i +q, 0) =  i (D-2d_i)$. 
The term $m_i^2/((k_i + q)^2 + m_i^2)$ should not be expanded in small $q$, 
since it then blows up on shell, where $k_i^2 = - m_i^2$.
Instead, as explained in Ref.~\cite{DiVecchia:2015jaq}, these terms should be 
kept through LSZ reduction, and taken on-shell yielding 
$m_i^2 /(2k_i \cdot q)$. It was then shown in Ref.~\cite{DiVecchia:2015jaq} 
that this procedure reproduces the correct mass-dependence of amplitude 
in the soft limit. For the sake of simplicity,  we will here, and throughout 
this work, neglect
such `Laurent' terms in the soft expansion and only focus on the `Taylor' terms.
 To be precise, we define 
\ea{
\tilde{F}_i^{\rm T}(k_i +q, m_i) = \tilde{F}_i(k_i +q, m_i) - 
\tilde{F}_i^{\rm L}(k_i +q, m_i)
\label{Ftaylor}
}
where $\tilde{F}_i^{\rm L}$ is the part of $\tilde{F}_i$ which on-shell has all 
the soft momentum poles of the form 
\ea{
\tilde{F}_i^{\rm L} \sim \sum_{n=1}^{\infty} \frac{L_n}{(k_i \cdot q)^n}
\label{Flaurent}
}
and thus $\tilde{F}_i^{\rm T}$ represents the part of $\tilde{F}_i$ which has a 
well-defined Taylor expansion on-shell.

We now have all the ingredients to write down the soft Ward identity. 
Considering for simplicity only the \emph{finite} parts of the soft limit as just 
described, i.e. neglecting parts belonging to the Laurent expansion, we get 
from \Eq{singlesoftWI-FT} for $\tilde{F}_i \to \tilde{F}_i^{\rm T}$:
\ea{
&i
f_\xi  \langle \txi(q) \tphi(k_1) \cdots \tphi(k_n) \rangle _{\rm amp} 
\nonumber \\
&
= i\sum_{i=1}^n
\left (
D-2d_i 
+ (d_i - D - (k_i+q) \cdot \partial_{k_i}  ) \right )
\langle  \tphi(k_1) \cdots \tphi(k_i+q)  \cdots \tphi(k_n) \rangle_{\rm amp} 
+\Ord(q)
\nonumber \\
&
= i\sum_{i=1}^n
\left (
-d_i  - k_i \cdot \partial_{k_i} \right )
\langle  \tphi(k_1) \cdots \tphi(k_i+q)  \cdots \tphi(k_n) \rangle_{\rm amp} 
+\Ord(q)
\label{singlesoftdilatonWI}
}
Using furthermore the commutation relation:
\ea{
\left [ \sum_{i=1}^n  k_i \cdot \partial_{k_i}   \, , 
\delta^{(D)}(\textstyle{\sum_i} k_i + q)
\right ] = - D  \delta^{(D)}(\textstyle{\sum_i} k_i + q)
}
which  according to \Eq{fg} defines the function \mbox{$\gamma_{\cal D}(k_i+q)= iD$},
we arrive at the soft theorem:
\ea{
 T_{n+1}(q;k_1, \ldots, \bar{k}_n) =
\frac{1}{f_\xi}\left [ D - \sum_{i=1}^n
\left ( d_i  + k_i \cdot \partial_{k_i} \right )
\right ] 
T_n(k_1, \ldots, \bar{k}_n ) + \Ord(q)
\label{leadingsoft}
}
This is a well-known expression dating back to works 
by G. Mack~\cite{Mack:1968zz}.
 It is worth observing that, due to the momentum conservation, the 
T-matrix, in \Eq{leadingsoft}, depends only on $(n-1)$ momenta. Therefore 
in the definition of dilatation operator, one of the momentum derivatives   
does not give any contribution  when evaluated on the amplitude. The   
\mbox{$\sum_{i=1}^n k_i\cdot \partial_{k_i}$ is thus   a sum on only $(n-1)$} 
momenta. 
This observation will be used in the sect.~\ref{dilatonamplitude} where the 
soft theorems will be  verified  on specific  amplitudes computed in  models 
with spontaneously broken conformal symmetry.  
   
The complete treatment given in simplified form here,
where all terms including those belonging to the soft Laurent expansion were 
taken into account, was performed in Ref.~\cite{DiVecchia:2015jaq}, where 
it was shown to also yield a soft factorizing theorem. The additional Laurent 
contributions automatically yield the terms that one can explicitly derive by
 Feynman diagram techniques, when noting that the dilaton couples linearly on 
the legs of massive external states. This was indeed the route taken in 
Ref.~\cite{Boels:2015pta}, but by our method it follows automatically from 
the Ward identity, as shown in Ref.~\cite{DiVecchia:2015jaq}.

\subsection{Soft Ward identity of special conformal transformations, $J_{\K, \lambda}^\mu$}
\label{singlesoftsct}
We specify in this section the general treatment to the
case of special conformal transformations with $J^\mu = J_{\K, \lambda}^\mu$
and $\Gamma_{\K, \lambda} = \K_\lambda$.
Following the notation of Sec.~\ref{CFT} we have
\seal{specialconformal0}{
\mathcal{K}_{\lambda, \varphi }(x) \varphi(x)
&=  i
\left( (2 x_\lambda x_{\nu} - \eta_{\lambda \nu} x^2 ) \partial^\nu 
+ 2\,d_\varphi\,x_\lambda 
+ 2i x^\nu \mathcal{S}_{ \nu \lambda }
\right)\varphi(x) 
\label{Kphi}\\[2mm]
\partial_\mu J_{\K,\lambda}^\mu = 
 &= 2\,x_{\lambda}\,T^{\mu}_{\phantom{\mu}\mu}
=2\,{f_\xi}\,x_{\lambda}(-\partial^2)\,\xi (x)
}
where $\varphi$ is any field and $d_\varphi$ is its scaling dimension.
The second expression defines $g_{\lambda} (x) = 2 f_\xi x_\lambda$.

To derive the Fourier transformed operators, we simply replace
every $x_\mu$ with a derivative $i \frac{\partial}{\partial k^\mu}$, while 
the derivative $\partial^\nu$ can be replaced with $ik^\nu$. Then after 
passing $k$-derivatives through $k^\nu$, one finds:
\eas{
\tilde{g}_\lambda (q) &= i 2f_\xi \partial_{q, \lambda}
\, ,  \\
\tilde{\K}_{\lambda, \varphi} (k) &=  
2 k_\nu \partial_k^\nu \partial_{k, \lambda}
- k_\lambda \partial_k^2 
- 2 ( d_\varphi -D) \partial_{k,\lambda}
+ 2 i\mathcal{S}_{\lambda \nu} \partial_k^\nu\label{Kmomentum}
}

To derive the commutation relations with the propagators, we need to specify 
the spin of the 
hard states to define the form of their propagator. Assuming for simplicity 
that the hard states are spin 0 scalar fields, we should neglect the spin operator.
Then it can be checked that the commutator with the scalar propagator reads:
\eas{
&\left [
\tilde{\K}_{\lambda, i}(k_i+q) \, , \
\Delta_i(k_i+q)
\right ] 
\\
&\qquad =
\Delta_i(k_i+q)
\left [
\frac{2( D-2d_i)   (k_i + q)_\lambda}{(k_i+q)^2 + m_i^2} 
\left (\frac{\left({ \frac{D}{2} - d_i } +1\right) m_i^2}{(k_i+q)^2 + m_i^2} 
\right )
\right ]
 \\
& \qquad \quad + 
\Delta_i(k_i+q)
\left [
-2( D-2d_i)\left (1- \frac{m_i^2}{(k_i+q)^2 + m_i^2} \right ) \partial_{k, \lambda} \right ]
}
where the first term is coming from the full action of $\tilde{\K}_{\lambda, i}$ on 
the propagator, while the second term arises due to the non-linearity 
of $\tilde{\K}_{\lambda, i}$, i.e. terms where one derivative hits 
the propagator and the other goes through. This expression defines 
the operator $\tilde{F}_{\lambda, i} (k_i +q, m_i)$, which due to the
 non-linearity of $\tilde{\K}_{\lambda, i}$ has a part which is not just 
a function, but a derivative operator. In reduced form:
\eas{
\tilde{F}_{\lambda, i} (k_i +q, m_i)
=
&- 2( D-2d_i) \partial_{k, \lambda}
 \\
&
+ \frac{2( D-2d_i)m_i^2}{(k_i+q)^2 + m_i^2}
 \left [
\left({ \frac{D}{2} - d_i } +1\right) \frac{(k_i + q)_\lambda}{(k_i+q)^2 + m_i^2}
+\partial_{k, \lambda}
\right ]
}
As in the previous section, we will here restrict our analysis to the part only belonging to the soft Taylor expansion, and refer to Ref.~\cite{DiVecchia:2015jaq} for the full treatment.
Thus, according to the definition in \Eq{Ftaylor}, we simply consider:
\ea{
\tilde{F}_{\lambda, i}  \to \tilde{F}^{\rm T}_{\lambda, i} (k_i +q, m_i)
=
2 (2d_i- D)\partial_{k, \lambda} 
\, , \quad
\text{for spinless $\varphi_i$ }
}
We note again that this is equivalent to the massless case, however, this restriction being more general.
By this prescription, we find from \Eq{singlesoftWI-FT} the following single-soft Ward identity:
\ea{
&-2f_\xi \partial_{q, \lambda} 
\langle \txi(q) \tphi(k_1) \cdots \tphi(k_n) \rangle _{\rm amp} 
 \\
&
= \sum_{i=1}^n
\left (2 (2d_i- D )\partial_{k_i, \lambda} + \tilde{\K}_{\lambda, i}(k_i+q)  \right )
\langle  \tphi(k_1) \cdots \tphi(k_i+q)  \cdots \tphi(k_n) \rangle_{\rm amp} 
+\Ord(q)
\nonumber \\
&
= \sum_{i=1}^n
\Big (2 d_i \partial_{k_i, \lambda} + 
2 (k_i +q)_\nu \partial_{k_i}^\nu \partial_{k_i, \lambda} - (k_{i}+ q)_{ \lambda} \partial_{k_i}^2
  \Big )
\langle  \tphi(k_1) \cdots \tphi(k_i+q)  \cdots \tphi(k_n) \rangle_{\rm amp} 
+\Ord(q)
\nonumber
}
It is useful to define the operator
\ea{
\hat{K}_{k_i}^\lambda = 
\frac{1}{2} k_i^\lambda \partial_{k_i}^2- (d_i  + k_i \cdot \partial_{k_i} ) \partial_{k_i}^\lambda \, ,
\label{hatK}
}
and then the single-soft Ward identity of special conformal transfomations {reads}:
\ea{
&f_\xi \partial_{q, \lambda} 
\langle \txi(q) \tphi(k_1) \cdots \tphi(k_n) \rangle _{\rm amp} 
= 
\sum_{i=1}^n
\hat{K}_{k_i + q, \lambda}
\langle  \tphi(k_1) \cdots \tphi(k_i+q)  \cdots \tphi(k_n) \rangle_{\rm amp} 
+\Ord(q)
\label{singlesoftWI-FT-sct}
}
Imposing LSZ reduction on this expression, and noting that the various operators involved all commute with the momentum conserving delta-function, this expression readily yields:
\ea{
f_\xi \partial_{q, \lambda} T_{n+1}(q; k_1, \ldots, \bar{k}_n)
= 
\sum_{i=1}^n
\hat{K}_{k_i + q, \lambda}T_n(k_1, \ldots, k_i + q, \ldots, \bar{k}_n)
+\Ord(q)
\label{subleadingsoft}
}
Since both sides of this expression should be evaluated for $q \sim 0$, it is clear that the left-hand side, when multiplied by $q_\lambda$ is the first order term in the Taylor expansion of $T_{n+1}$ around $q = 0$. Thus:
\eas{
&T_{n+1} (q, k_1, \ldots , \bar{k}_n) =  T_{n+1} (0, k_1, \ldots , \bar{k}_n) + 
 q^\mu \frac{\partial}{\partial q^\mu} T_{n+1} (0, k_1, \ldots , \bar{k}_n) + \Ord(q^2)
\\
&=
\frac{1}{f_\xi} \left [ D - \sum_{i=1}^n
\left ( d_i  + k_i \cdot \partial_{k_i} \right )
+ q^\lambda \sum_{i=1}^n \hat{K}_{k_i, \lambda}
\right ] 
T_n(k_1, \ldots, \bar{k}_n ) + \Ord(q^2)
}
where we used \Eq{leadingsoft} for the leading term in the expansion 
and  \Eq{subleadingsoft} for the subleading term. For the Laurent terms, 
{the order of soft limit and on-shell limit is subtle and must be performed 
with care},  nevertheless it is possible to show that one can derive the soft 
theorem through subleading order~\cite{DiVecchia:2015jaq} following the 
same procedure, including all the correct terms of the Laurent expansion.

\section{Double-Soft Ward Identity and double-soft dilaton theorem}
\label{doublesoftgeneral}
In this section we apply the same current algebra procedures as defined and 
performed in the preceeding section, but with the complication of 
inserting two 
currents in the matrix element of $T^\ast$-ordered product of operators.
This leads to new soft Ward identities as well as a new double-soft theorem for
 the dilaton.

We consider the forementioned matrix element and take space-time derivatives on the space-time variables of the two currents. In addition to the single-soft assumption of \Eq{assumption}, we similarly assume
\ea{
\int d^D y \,\,{\rm e}^{-ik y} \int d^D x  \,\,{\rm e}^{-i q x} \partial_\nu^y 
\partial_\mu^x T^\ast \langle 0| J_1^\mu (x) J_2^\nu (y) \phi (x_1) \cdots
 \phi (x_n) |0 \rangle =
 0 + \Ord(k_\nu q_\mu)
 \label{assumption2}
}
This follows from taking the Fourier transform of the derivatives, and assuming that the correlation function has no poles in the momentum variables of the currents.

Considering instead the action of the derivatives on the matrix element we find:
\ea{
\begin{split}
\int &d^D y \,\,{\rm e}^{-ik y} \int d^D x  \,\,{\rm e}^{-i q x} \partial_\nu^y 
\partial_\mu^x T^\ast \langle 0| J_1^\mu (x) J_2^\nu (y) \phi (x_1) \cdots
 \phi (x_n) |0 \rangle
  \\
 =& \int d^D y \,\,{\rm e}^{-ik y}   \partial_\nu^y  \left[ 
\int d^D x  \,\,{\rm e}^{-i q x} T^\ast
\langle 0| (\partial_\mu J_1^\mu (x)) J_2^\nu (y) \phi (x_1) \cdots
 \phi (x_n) |0 \rangle  \right.
  \\
& + {\rm e}^{-i q y} T^\ast \langle 0| [ Q_1, J_2^\nu (y) ] 
\phi (x_1) \cdots
 \phi (x_n) |0 \rangle 
  \\
& \left.  + \sum_{i=1}^n  {\rm e}^{- iq x_i}
T^\ast \langle 0| J_2^\nu (y) \phi (x_1 )
\cdots [ Q_1, \phi (x_i)] \cdots \phi (x_n ) |0\rangle  \right]
 \\
=
 &
 \int d^D y \,\,{\rm e}^{-ik y}     
\int d^D x  \,\,{\rm e}^{-i q x} T
\langle 0| (\partial_\mu J_1^\mu (x))  (\partial_\nu^y  
 J_2^\nu (y) )\phi (x_1) \cdots
 \phi (x_n) |0 \rangle   
  \\
& + \int d^D x  \,\,{\rm e}^{-i (q+k) x} T^\ast \langle 0|\,\, [Q_2, \partial_\mu
 J_1^\mu (x) ] \, \, \phi (x_1) \cdots \phi (x_n)|0\rangle
 \\
& + \sum_{i=1}^n \int d^D x  \,\, {\rm e}^{-i (q x +k x_i)} T^\ast \langle 0|
(\partial_\mu J_1^\mu (x)) \phi (x_1 ) \cdots [Q_2, \phi (x_i )] \cdots
\phi (x_n ) |0 \rangle 
\\
&
 +  \int d^D y \,\,{\rm e}^{-ik y}   \partial_\nu^y 
\left ( {\rm e}^{-i q y} T^\ast \langle 0| [ Q_1, J_2^\nu (y) ] 
\phi (x_1) \cdots
 \phi (x_n) |0 \rangle \right )
  \\
& + \sum_{i=1}^n \int d^D y  \,\, {\rm e}^{-i (k y  +q x_i)} T^\ast \langle 0|
(\partial_\nu J_2^\nu (y)) \phi (x_1 ) \cdots [Q_1, \phi (x_i )] \cdots
\phi (x_n ) |0 \rangle 
 \\
& + \sum_{i \neq j} {\rm e}^{- iq x_i} {\rm e}^{- ik x_j}
T^\ast \langle 0| \phi (x_1) \cdots [Q_2, \phi (x_j ) ]\cdots
[Q_1, \phi (x_i )] \cdots \phi (x_n ) |0 \rangle  
 \\
& + \sum_{i=1}^{n} {\rm e}^{- i(q+k) x_i} T^\ast\langle 0| \phi (x_1 ) \cdots
[Q_2, [ Q_1, \phi (x_i )]] \cdots \phi (x_n ) |0\rangle
\end{split}
\label{GeneralWI}
}
where we are assuming $x\neq y$.
This expression can be further reduced by using the single-soft Ward identity 
in \Eq{singlesoftWI},
as well as the identities in \Eq{gJ} and \Eq{QMN}.
Let us remark that the left-hand side of this Ward identity is manifestly  
symmetric under $q \leftrightarrow k$ and $J_1 \leftrightarrow J_2$. This 
means that our end result for the right-hand side must as well possesses
this 
symmetry. For simplicity, we impose this 
at the end, but in principle the above 
expression could be already symmetrized.

The left-hand side of \Eq{GeneralWI} is by \Eq{assumption2} zero up to $\Ord(k^\nu q^\mu)$.
The first term on the right-hand side 
 can by \Eq{gJ} be reduced to:
\eas{
&\int d^D y \,\,{\rm e}^{-ik y}  
\int d^D x  \,\,{\rm e}^{-i q x} T^\ast
\langle 0| (\partial_\mu J_1^\mu (x)) (\partial_\nu J_2^\nu (y) ) \phi (x_1) \cdots
 \phi (x_n) |0 \rangle 
  \\
 &=
 \int d^D y \,\,{\rm e}^{-ik y}   g_2 (y)  
\int d^D x \,\,{\rm e}^{-i q x} g_1(x) T^\ast
(-\partial_x^2) (-\partial_y^2)\langle 0| \xi(x) \xi(y) \phi (x_1) \cdots
 \phi (x_n) |0 \rangle 
 \\
 &=
 \tilde{g}_1(q) \tilde{g}_2(k) \left (k^2 q^2 
 \langle \txi(q) \txi(k) \phi (x_1) \cdots
 \phi (x_n)  \rangle  \right )
}
Performing the Fourier transform of the remaining fields gives:
\eas{
 &\tilde{g}_1(q) \tilde{g}_2(k)\, k^2\, q^2
\int \prod_{j=1}^n\Big[\frac{d^D k_i}{(2\pi)^D}e^{-ik_jx_j}\big] \langle \txi(q) \txi(k) \phi (x_1) \cdots
 \phi (x_n)  \rangle\\
 &=- \prod_{i=1}^n 
 \Delta_i(k_i)
 \tilde{g}_1(q) \tilde{g}_2(k)  
 \langle \txi(q) \txi(k) \tphi (k_1) \cdots
 \tphi (k_n)  \rangle_{\rm amp}
}
where
the correlation function on the right-hand side is amputated, and $\Delta_i$ are the two-point correlation functions of the fields $\phi_i$, defined in \Eq{Deltai}.

The second term can be simplified as follows
\eas{
&\int d^D x  \,\,{\rm e}^{-i (q+k) x} T^\ast \langle 0|\, [Q_2, \partial_\mu
 J_1^\mu (x) ] \, \, \phi (x_1) \cdots \phi (x_n)|0\rangle 
  \\
 &=
 \int d^D x  \,\,{\rm e}^{-i (q+k) x}  T^\ast \langle 0|\, [Q_2, g_1(x)(-\partial^2)\xi (x) ] \, \, \phi (x_1) \cdots \phi (x_n)|0\rangle 
 \\
 &=
 \int d^D x  \,\,{\rm e}^{-i (q+k) x} {\Gamma}_{2, g_1 \partial^2 \xi}(x)  {g}_1(x) T^\ast \langle 0|\,\, (-\partial^2)\xi (x) \, \phi (x_1) \cdots \phi (x_n)|0\rangle 
 \\
 &=
 \tilde{\Gamma}_{2, g_1 \partial^2 \xi}(q+k)
  \tilde{g}_1(q+k) 
\left ((q+k)^2
 \langle \txi(q+k)  \phi (x_1) \cdots
 \phi (x_n)  \rangle  \right )
}
where $\tilde{\Gamma}_{2g_1 \partial^2 
\xi}(q+k)$ is the Fourier transform of the generator of infinitesimal 
transformations related to $Q_2$ and $g_1 \partial^2 \xi$, as defined in 
\Eq{deltaphi}.
Again, performing the Fourier transform of the remaining fields gives:
\eas{
&\tilde{\Gamma}_{2, g_1 \partial^2 \xi}(q+k)
  \tilde{g}_1(q+k) 
(q+k)^2\int \prod_{j=1}^n\Big[\frac{d^D k_i}{(2\pi)^D}e^{-ik_jx_j}\big]\langle \txi(q+k)  \phi (x_1) \cdots
 \phi (x_n)  \rangle 
\\
 &=- i 
 \prod_{i=1}^n  \Delta_i(k_i)
 \tilde{\Gamma}_{2, g_1 \partial^2 \xi}(q+k)
  \tilde{g}_1(q+k) 
 \langle \txi(q+k)  \tphi (k_1) \cdots
 \tphi (k_n)  \rangle_{\rm amp}
}
We cannot reduce this expression further, since we need to know the explicit 
form of the operator $ \tilde{\Gamma}_{2, g_1 \partial^2 \xi}$, which acts 
on both $\tilde{g}_1$ and the $(n+1)$-point amputated correlation function, 
involving the dilaton. We will later see that when one of the associated
currents is the dilatation current, this expression can be further reduced by 
using the single-soft theorem of the previous section.

The third term on the right-hand side of \Eq{GeneralWI} can similarly be reduced to:
\eas{
&\sum_{i=1}^n \int d^D x  \,\, {\rm e}^{-i (q x +k x_i)} T^\ast \langle 0|
(\partial_\mu J_1^\mu (x)) \phi (x_1 ) \cdots [Q_2, \phi (x_i )] \cdots
\phi (x_n ) |0 \rangle 
 \\
&=
\sum_{i=1}^n  {\rm e}^{-i k x_i}
\tilde{g}_1(q)
\int d^D x  \,\, {\rm e}^{-i q x} T^\ast \langle 0|
(-\partial^2)\xi(x) \phi (x_1 ) \cdots 
{\Gamma}_{2, \phi_i}(x_i) \phi(x_i)
\cdots
\phi (x_n ) |0 \rangle 
 \\
&=
\sum_{i=1}^n  {\rm e}^{-i k x_i} 
\tilde{g}_1(q) \left (q^2  \langle \txi(q) \phi (x_1 ) \cdots 
{\Gamma}_{2, \phi_i}(x_i) \phi(x_i)
\cdots
\phi (x_n ) \rangle  \right )
 }
This expression can be further reduced by making use of the
singe-soft Ward identity given in \Eq{singlesoftWI2}, after Fourier transforming also the $x_i$ variables. Thus, the previous expression transforms to
\eas{
& 
 \sum_{i=1}^n
\tilde{g}_1(q) \left (q^2  \langle \txi(q) \tphi (k_1 ) \cdots 
\tilde{\Gamma}_{2, \phi_i}(k_i + k) \tphi(k_i + k)
\cdots
\tphi (k_n ) \rangle  \right )
 \\
&=
- \sum_{i=1}^n
\sum_{j\neq i}^n
\tilde{\Gamma}_{1, \phi_j}(k_j+q) \tilde{\Gamma}_{2, \phi_i}(k_i+k)
\langle  \tphi(k_1) \cdots \tphi(k_j+q) \cdots  \tphi(k_i + k) \cdots \tphi(k_n) \rangle
 \\
& \quad - \sum_{i=1}^n \tilde{\Gamma}_{1, {} \phi_i}(k_i+k+q) \tilde{\Gamma}_{2, \phi_i}(k_i+k+q)
\langle  \tphi(k_1) \cdots  \tphi(k_i + k+q) \cdots \tphi(k_n) \rangle + \Ord(q)
}
 where care was taken on using the soft Ward identity for $j = i$.
 We may now amputate the correlation function, which can be expressed using the
 definition for $\tilde{F}$ in \Eq{FJ}
 \ea{
 & \sum_{i=1}^n
\tilde{g}_1(q) \left (q^2  \langle \txi(q) \tphi (k_1 ) \cdots 
\tilde{\Gamma}_{2, \phi_i}(k_i + k) \tphi(k_i + k)
\cdots
\tphi (k_n ) \rangle  \right ) \nn
 \\
&
=
- \prod_{l=1}^n \Delta_l(k_l)
\sum_{i=1}^n
\left (
\tilde{F}_{2, \phi_i }(k_i + k, m_i)
+
\tilde{\Gamma}_{2, \phi_i}(k_i + k)
\right )
\sum_{j \neq 1}^n 
\left (\tilde{F}_{1, \phi_j}(k_j+q, m_j) + \tilde{\Gamma}_{1, \phi_j}(k_j+q)  \right )
\nonumber \\
& \qquad \times
\langle  \tphi(k_1) \cdots \tphi(k_i+k) \cdots  \tphi(k_j + q)  \cdots \tphi(k_n) \rangle_{\rm amp} 
\nonumber \\
&\quad - \prod_{l=1}^n \Delta_l(k_l)
\sum_{i=1}^n
\left (\tilde{F}_{1, {}\phi_i}(k_i+k + q, m_i) + \tilde{\Gamma}_{1, {} \phi_i}(k_i + k+q)  \right )
\nonumber \\
&\qquad \times
\left (
\tilde{F}_{2, \phi_i }(k_i + k+q, m_i)
+
\tilde{\Gamma}_{2, \phi_i}(k_i + k+q)
\right )
\nonumber \\
& \qquad \times
\langle  \tphi(k_1) \cdots \tphi(k_i+k+q)   \cdots \tphi(k_n) \rangle_{\rm amp} 
+\Ord(q, k)
\label{3rd}
}
where we took the limit $k, q \to 0$ in the propagators $\Delta_l$.

In the fourth term of the right-hand side of \Eq{GeneralWI} we did not act with the derivative $\partial_\nu^y$, because we instead Fourier transform it to show that the term is of $\Ord(k)$ by assumption:
 \eas{
 & \int d^D y \,\,{\rm e}^{-ik y}   \partial_\nu^y 
\left ( {\rm e}^{-i q y} T^\ast \langle 0| [ Q_1, J_2^\nu (y) ] 
\phi (x_1) \cdots
 \phi (x_n) |0 \rangle \right )
 \\
 &
 =
 i k_\nu 
  \int d^D y \,\,{\rm e}^{-i(k+q) y}
 {\Gamma}_1(y) 
 T^\ast \langle 0| J_2^\nu (y) 
\phi (x_1) \cdots
 \phi (x_n) |0 \rangle
\\
 &
 =
 i k_\nu \tilde{\Gamma}_1 (k+q)
  \int d^D y \,\,{\rm e}^{-i(k+q) y}
 T^\ast \langle 0| J_2^\nu (y) 
\phi (x_1) \cdots
 \phi (x_n) |0 \rangle
 \\
 & 
 = 0 + \Ord(k)
 }
where the last line follows from  Eq.~\eqref{assumption} (as well as assuming no pole in $\tilde{\Gamma}_1$).

The fifth term is equivalent to the third term, but with the symmetry indices interchanged $1 \leftrightarrow 2$ and the soft-momenta likewise interchanged $q \leftrightarrow k$. Thus the fifth term gives:
\ea{
&
 \int \prod_{j=1}^n \Big[d^D x_j \,\, {\rm e}^{-i k_j x_j}\Big]
\sum_{i=1}^n \int d^D y  \,\, {\rm e}^{-i (k y  +q x_i)} T^\ast \langle 0|
(\partial_\nu J_2^\nu (y)) \phi (x_1 ) \cdots [Q_1, \phi (x_i )] \cdots
\phi (x_n ) |0 \rangle 
 \nonumber \\
 &=
- \prod_{l=1}^n \Delta_l(k_l)
\sum_{i=1}^n
\left (
\tilde{F}_{2, \phi_i }(k_i + k, m_i)
+
\tilde{\Gamma}_{2, \phi_i}(k_i + k)
\right )
\sum_{j \neq 1}^n 
\left (\tilde{F}_{1, \phi_j}(k_j+q, m_j) + \tilde{\Gamma}_{1, \phi_j}(k_j+q)  \right )
\nonumber \\
& \qquad \times
\langle  \tphi(k_1) \cdots \tphi(k_i+k) \cdots  \tphi(k_j + q)  \cdots \tphi(k_n) \rangle_{\rm amp} 
\nonumber \\
&\quad - \prod_{l=1}^n \Delta_l(k_l)
\sum_{i=1}^n
\left (\tilde{F}_{2, {}\phi_i}(k_i+k + q, m_i) + \tilde{\Gamma}_{2, {} \phi_i}(k_i + k+q)  \right )
\nonumber \\
&\qquad \times
\left (
\tilde{F}_{1, \phi_i }(k_i + k+q, m_i)
+
\tilde{\Gamma}_{1, \phi_i}(k_i + k+q)
\right )
\nonumber \\
& \qquad \times
\langle  \tphi(k_1) \cdots \tphi(k_i+k+q)   \cdots \tphi(k_n) \rangle_{\rm amp} 
+\Ord(q, k)
\label{5th}
}
The terms with the double sum, where $j\neq i$, are the same as before since the operators here commute. 
The operators in the single sum, on the other hand, do not commute. Instead 
these terms,
together with the similar ones in \Eq{3rd}, add up to ensure the symmetry $q \leftrightarrow k$ and $J_1 \leftrightarrow J_2$, which is manifest on the left-hand side of the Ward identity.

The sixth term leads to
\eas{
&\sum_{i \neq j} {\rm e}^{- iq x_i} {\rm e}^{- ik x_j}
T^\ast \langle 0| \phi (x_1) \cdots [Q_2, \phi (x_j ) ]\cdots
[Q_1, \phi (x_i )] \cdots \phi (x_n ) |0 \rangle 
\\
&=
\sum_{i \neq j} {\rm e}^{- ik x_j} {\Gamma}_{2,\phi_j}(x_j)   {\rm e}^{- iq x_i} {\Gamma}_{1,\phi_i} (x_i)
T^\ast \langle 0| \phi (x_1) \cdots\phi (x_n ) |0 \rangle 
}
It is easy to see that by taking the Fourier transform and amputating the 
correlation function, 
this expression exactly cancels the similar terms with double sums in, either 
the third expression in  \Eq{3rd} or the fifth expression in \Eq{5th}.

Finally, for the seventh term we {make use 
of} the Jacobi identity:
\ea{
[Q_2, [ Q_1, \phi_i ]] = [[Q_2, Q_1], \phi_i] + [Q_1, [Q_2, \phi_i]]
}
As mentioned earlier, the left-hand side of the Ward identity is manifestly symmetric under $q \leftrightarrow k$, $J_1 \leftrightarrow J_2$. To ensure the symmetry on the right-hand side we should symmetrize the seventh term. This symmetrization gets rid of the commutator $[Q_2, Q_1]$ above and sends:
\ea{
[Q_2, [ Q_1, \phi_i ]] \to \frac{1}{2} \left (\Gamma_{2, {} \phi_i} \Gamma_{1, \phi_i} + \Gamma_{1, {} \phi_i} \Gamma_{2, \phi_i} \right ) \phi_i
}
Thus the seventh term by symmetrization is the sum of the two terms
\eas{
&\frac{1}{2}\sum_{i=1}^{n} {\rm e}^{- i(q+k) x_i} T^\ast\langle 0| \phi (x_1 ) \cdots
[Q_2, [ Q_1, \phi (x_i )]] \cdots \phi (x_n ) |0\rangle +(1\leftrightarrow 2)
 \\
&= \sum_{i=1}^{n} {\rm e}^{- i(q+k) x_i}\frac{1}{2} \left (\Gamma_{2, {} \phi_i} \Gamma_{1, \phi_i} + \Gamma_{1, {} \phi_i} \Gamma_{2, \phi_i} \right )
T^\ast\langle 0| \phi (x_1 ) \cdots \phi (x_i )\cdots \phi (x_n ) |0\rangle
}
It is readily seen that after Fourier transforming and amputating, this expression 
cancels one half of the similar terms in \Eq{3rd} and \Eq{5th}.

Finally, taking into account the symmetrization, we can express the full double-soft Ward identity on amputated correlation functions in momentum space:
 \ea{
 &\tilde{g}_1(q) \tilde{g}_2(k)  
 \langle \txi(q) \txi(k) \tphi (k_1) \cdots
 \tphi (k_n)  \rangle_{\rm amp}\mid_{q,k \sim 0}
\nonumber \\[2mm]
&
=
- \frac{i}{2} \left [
 \tilde{\Gamma}_{2, g_1 \partial^2 \xi}(q+k)
  \tilde{g}_1(q+k) 
  +
   \tilde{\Gamma}_{1, g_2 \partial^2 \xi}(q+k)
  \tilde{g}_2(q+k) 
  \right ]
 \langle \txi(q+k)  \tphi (k_1) \cdots
 \tphi (k_n)  \rangle_{\rm amp}
  \nonumber \\
 &
 \quad - 
\Big [ \sum_{i=1}^n \left ( \tilde{F}_{1, \phi_i} (k_i + k , m_i) + \tilde{\Gamma}_{1, \phi_i}(k_i + k) \right )
\sum_{j \neq i}^n 
\left (\tilde{F}_{2, \phi_j}(k_j + q, m_j) + \tilde{\Gamma}_{2, \phi_j}(k_j + q)  \right )
\nonumber \\
&+\frac{1}{2}\sum_{i=1}^n \left ( \tilde{F}_{1, {}\phi_i} (k_i + k + q , m_i) + \tilde{\Gamma}_{1, {} \phi_i}(k_i + k + q) \right )
\left (\tilde{F}_{2, \phi_i}(k_i+ k + q, m_i) + \tilde{\Gamma}_{2, \phi_i}(k_i+ k + q)  \right )
\nonumber \\
&+\frac{1}{2}\sum_{i=1}^n 
\left (\tilde{F}_{2, {} \phi_i}(k_i+k+q, m_i) + \tilde{\Gamma}_{2, {} \phi_i}(k_i + k + q)  \right )
\left ( \tilde{F}_{1,  \phi_i} (k_i + k + q , m_i) + \tilde{\Gamma}_{1, \phi_i}(k_i + k + q) \right )
\Bigg ]
\nonumber
\\
&\qquad \quad \times
\langle  \tphi(k_1) \cdots \tphi(k_i+k) \cdots  \tphi(k_j + q) \cdots  \tphi(k_n) \rangle_{\rm amp} 
+\Ord(q,k)
\label{doublesoftWI}
 }
 where in the last correlator it is implicitly assumed that for the single-sum expressions 
one should understand $ \tphi(k_i+k) \cdots  \tphi(k_i + q)  \sim  \tphi(k_i+k+q) $.

In the case of massless hard states, the limit $q, k \to 0$ may be well-behaved.
If that is so, and  if  furthermore $[\Gamma_1, \Gamma_2] = 0$, then
the above expression simplifies to:
 \ea{
&\tilde{g}_1(q) \tilde{g}_2(k)  
 \langle \txi(q) \txi(k) \tphi (k_1) \cdots
 \tphi (k_n)  \rangle_{\rm amp}\mid_{q,k \sim 0}
\label{masslessdoublesoftWI}
 \\[2mm]
&
=
- \frac{i}{2} \left [
 \tilde{\Gamma}_{2, g_1 \partial^2 \xi}(q+k)
  \tilde{g}_1(q+k) 
  +
   \tilde{\Gamma}_{1, g_2 \partial^2 \xi}(q+k)
  \tilde{g}_2(q+k) 
  \right ]
 \langle \txi(q+k)  \tphi (k_1) \cdots
 \tphi (k_n)  \rangle_{\rm amp}
  \nonumber \\
 &
 \quad - 
 \sum_{i=1}^n \left ( \tilde{F}_{1, \phi_i} (k_i) + \tilde{\Gamma}_{1, \phi_i}(k_i) \right )
\sum_{j = i}^n 
\left (\tilde{F}_{2, \phi_j}(k_j) + \tilde{\Gamma}_{2, \phi_j}(k_j)  \right )
\langle  \tphi(k_1) \cdots  \tphi(k_n) \rangle_{\rm amp} 
+\Ord(q,k)
\nonumber
 }

 \subsection{Double-soft Ward identity of two dilatation currents}
\label{doublesoftDD}
We specialize the previous analysis to the case of two dilatation current 
insertions in the matrix element.
 Following the definitions and expressions in Sec.~\ref{singlesoftdilatation},
we have
\eas{
\tilde{g}_1 = \tilde{g}_2 &= f_\xi \, , 
\qquad 
\tilde{\D}_{i}(k) = i (d_i - D - k \cdot \partial_k ) \, , 
\\[2mm]
\tilde{F}_i(k_i +q, m_i) &=
i (D-2d_i)\left ( 1- \frac{m_i^2}{(k_i+q)^2 + m_i^2} \right ) \, 
\\[2mm]
f_i (k_i +q, m_i) &= 0 \, ,
\qquad  
\gamma_i (k_i +q)  = i D \ .
}
We will in this work only focus on the parts of the double-soft Ward identities 
belonging to the Taylor expansion in the soft momenta, as described and
 prescribed 
in \Eq{Ftaylor}. In this case, this is equivalent to setting 
$\tilde{F}_i \to \tilde{F}_i^{\rm T} =\tilde{F}_i (k_i +q, 0)$.  
Due to this restriction and since $[\D, \D ]=0$, 
we need only to consider the simpler form of the double-soft Ward identity 
in Eq.~\eqref{masslessdoublesoftWI}.

Let us first consider the first term on the right-hand side of \Eq{masslessdoublesoftWI}, which under the above specifications takes the form:
\eas{
&- i \tilde{\Gamma}_{2, g_1 \partial^2 \xi}(q+k)
  \tilde{g}_1(q+k) 
 \langle \txi(q+k)  \tphi (k_1) \cdots
 \tphi (k_n)  \rangle_{\rm amp} 
 \\
 &=
 (d_{\partial^2\xi} - D - (k+q) \cdot \partial_{k+q}) f_\xi  \langle \txi(q+k)  \tphi (k_1) \cdots
 \tphi (k_n)  \rangle_{\rm amp} \label{5.17}
 }
 Now using the single-soft Ward identities given in Eq.~\eqref{singlesoftdilatonWI} it follows that the right-hand side of \Eq{5.17} is equal to:
 \ea{
  (d_{\partial^2\xi} - D)   \sum_{i=1}^n (-d_i - k_i\cdot \partial_{k_i} ) \langle   \tphi (k_1) \cdots
  \tphi(k_i + k+ q) \cdots 
 \tphi (k_n)  \rangle_{\rm amp}  + \Ord(k+q)
 }
Then it is straightforward to write the full expression 
for \Eq{masslessdoublesoftWI}:
\ea{
 &f_\xi^2
 \langle \txi(q) \txi(k) \tphi (k_1) \cdots
 \tphi (k_n)  \rangle_{\rm amp}
 \\
&
=
 (d_{\partial^2\xi} - D)   \sum_{i=1}^n (-d_i - k_i\cdot \partial_{k_i} ) 
  \langle   \tphi (k_1) \cdots
  \tphi(k_i + k+ q) \cdots 
 \tphi (k_n)  \rangle_{\rm amp} 
\nonumber \\
& \quad
 - 
 \sum_{i=1}^n \left ( i(D-2d_i) + i (d_i - D - k_i\cdot \partial_{k_i} ) \right )
\sum_{j =1}^n 
\left (i(D-2d_j)+ i (d_j - D - k_j\cdot \partial_{k_j} ) \right )
\nonumber \\
&\qquad  \times
\langle  \tphi(k_1) \cdots  \tphi(k_n) \rangle_{\rm amp} 
+\Ord(q,k)
\nonumber \\[2mm]
&
=
 \sum_{i=1}^n \left ( -d_i  - k_i\cdot \partial_{k_i} \right )
 \left [
(d_{\partial^2\xi} - D)
 +
\sum_{j =1}^n 
\left (-d_j  - k_j\cdot \partial_{k_j}  \right )
\right]
\langle  \tphi(k_1) \cdots  \tphi(k_n) \rangle_{\rm amp} 
+\Ord(q,k)
\nonumber
}

Since $\partial^2 \xi$ is the second descendant of the primary field, $\xi$, the dilaton, 
it follows that
\ea{
d_{\partial^2 \xi} = 2 + d_\xi 
= D-d_\xi
}
Thus
\ea{
 &f_\xi^2
 \langle \txi(q) \txi(k) \tphi (k_1) \cdots
 \tphi (k_n)  \rangle_{\rm amp}
\\
&
=
 \sum_{i=1}^n \left ( -d_i  - k_i\cdot \partial_{k_i} \right )
 \left (
-d_{\xi} 
 +
\sum_{j =1}^n 
\left (-d_j  - k_j\cdot \partial_{k_j}  \right )
\right )
\langle  \tphi(k_1) \cdots  \tphi(k_n) \rangle_{\rm amp} 
+\Ord(q,k)
\nonumber 
}
This expression is nothing but two consecutive applications of the single-soft Ward identity, where in the first application, one of the dilatons is taken to be hard. This shows that there is no difference at leading order between the two limits: $q \sim k \ll k_i$ and $q \ll k \ll k_i$.

We can go on and express this in terms of amplitudes by performing the LSZ reduction.
This gives us the double-soft theorem:
\ea{
\begin{split}
f_\xi^2 T_{n+2}(q,k, k_1, \ldots, \bar{k}_n)
= 
&
\Bigg [
D-d_\xi
 - \sum_{j =1}^n 
\left (d_j  + k_j\cdot \partial_{k_j}  \right )
\Bigg]
\left [D -  \sum_{i=1}^n \left ( d_i  + k_i\cdot \partial_{k_i} \right ) \right ]
 \\
&\times
T_{n}(k_1, \ldots, \bar{k}_n)
+\Ord(q,k)
\end{split}
\label{leadingdoublesoft}
}
This is again nothing but two single-soft dilaton theorems applied consecutively.
Thus there is no distinction between two soft dilatons emitted consecutively with two soft dilatons emitted simultaneously.
The bar on $k_n$ means that we keep one of the hard momenta, say $k_n$, fixed by momentum conservation, as in \Eq{kn}.

 In the case where all fields have free scalar field dimension $d_i = d_\xi = d = (D-2)/2$,
then
 \eas{
f_\xi^2 T_{n+2}(q,k, k_1, \ldots, \bar{k}_n)
= 
&
\Bigg [
D - (n+1)d
 - \sum_{j =1}^n 
  k_j\cdot \partial_{k_j}
\Bigg]
\left [D -  nd - \sum_{i=1}^n  k_i\cdot \partial_{k_i}  \right ]
 \\
&\times
T_{n}(k_1, \ldots, \bar{k}_n)
+\Ord(q,k)
}
 \Eq{leadingdoublesoft} is, however, more general, since the hard states can be interacting fields carying anomalous dimension. We can parametrize this by denoting $d_i = d + \eta_i$, while still $d_\xi = d$, then:
 \eas{
f_\xi^2 T_{n+2}(q,k, k_1, \ldots, \bar{k}_n)
= 
&
\left [
D - (n+1)d
 - \sum_{j =1}^n (\eta_j + 
  k_j\cdot \partial_{k_j} )
\right]
 \\
\times &\left [D -  nd - \sum_{i=1}^n (\eta_i +  k_i\cdot \partial_{k_i} ) \right ]
T_{n}(k_1, \ldots, \bar{k}_n)
+\Ord(q,k)
}
{where $\eta_i$ are the anomalous dimensions of the scalar fields $\phi_i$.}

\subsection{Double-soft Ward identity of the two currents, $J_\D^\mu$ and $J_{\K, \lambda}^\mu$}
\label{doublesoftQ1Q2}

We consider the double-soft Ward identity in \Eq{doublesoftWI}, following 
insertions of a dilatation current, $J_\D^\mu$, and a special conformal
 transformation current, $J_{\K, \lambda}^\mu$, in the matrix
element \Eq{matrixelements}.
 Following the definitions and expressions in 
Sect.~\ref{CFT},~\ref{singlesoftdilatation} and \ref{singlesoftsct},
 as well as the restriction described at \Eq{Ftaylor},
we take
 \ea{
 \begin{split}
\tilde{g}_1 = {f_\xi} \, , \qquad&
  \tilde{g}_{2, \lambda}(k) = i 2 {f_\xi} \partial_{k, \lambda} \, , \\
\tilde{\D}_{i}(k) = i (d_i - D - k \cdot \partial_k ) \, , \qquad&
\tilde{\K}_{\lambda, i} (k) =  
2 k^\nu \partial_{k,\nu} \partial_{k, \lambda}
- k_\lambda \partial_k^2
- 2 (d_i-D) \partial_{k,\lambda}
\\
 \tilde{F}_{1,\phi_i}^{\rm T} (k_i +q, m_i) =
\, i (D- 2d_i)
\, , \qquad
&\tilde{F}_{2,\phi_i}^{{\rm T}, \lambda} (k_i +q, m_i) = 
-2(D-2d_i) \partial_{k}^{\lambda}
\, , \\
f_{1,\phi_i} (k_i +q, m_i) = 0 \, , \qquad &
\gamma_{1, \phi_1} (k_i +q)  = D \, , \\
f_{2,\phi_i}^\lambda (k_i +q, m_i) = 0 \, , \qquad &
\gamma_{2, \phi_i}^\lambda (k_i +q)  = 0 \, , 
\end{split}
\label{DK}
}
For consistency it can be checked that:
\ea{
[\tilde{\D} , \tilde{\K}_\lambda ] = i \tilde{\K}_\lambda
}
This is in fact true for any value of $d_i$ and thus this term in $\tilde{\K}_\lambda$ can take any prefactor and still preserve the commutation relation above.

Let us consider the first line on the right-hand side of \Eq{doublesoftWI}, reading:
\ea{
- \frac{i}{2}& \left [
 \tilde{\Gamma}_{2, g_1 \partial^2 \xi}(q+k)
  \tilde{g}_1(q+k) 
  +
   \tilde{\Gamma}_{1, g_2 \partial^2 \xi}(q+k)
  \tilde{g}_2(q+k) 
  \right ]
 \langle \txi(q+k)  \tphi (k_1) \cdots
 \tphi (k_n)  \rangle_{\rm amp}
 \nonumber \\[2mm]
 &=
 - \frac{i}{2} \left [
 f_\xi \K_{\lambda, \partial^2 \xi} (q+k)
 + i 2 f_\xi \D_{x\partial^2 \xi} (q+k) \partial_{k+q, \lambda}
 \right ]
 \langle \txi(q+k)  \tphi (k_1) \cdots
 \tphi (k_n)  \rangle_{\rm amp}
 \nonumber \\[2mm]
 &=
 i f_\xi 
 \left [
 d_{\partial^2\xi } 
 + d_{x\partial^2 \xi} 
 -2D
 \right ]
  \partial_{k+q, \lambda}
  \langle \txi(q+k)  \tphi (k_1) \cdots
 \tphi (k_n)  \rangle_{\rm amp} + \Ord(k+q)
 \label{lhsDK}
}
The last expression can be further reduced by making use of the single-soft Ward identity for special conformal transformations, given in \Eq{singlesoftWI-FT-sct}, getting
\ea{
= i (d_{\partial^2 \xi}+d_{x\partial^2 \xi} - 2D) 
&\sum_{i=1}^n
\hat{K}_{k_i, \lambda}
\langle  \tphi(k_1) \cdots  \tphi(k_n) \rangle_{\rm amp} 
+ \Ord(k+q)
}
where $\hat{K}_{k_i, \lambda}$ was defined in \Eq{hatK}, and differs from $\tilde{\K}_{i,\lambda}$ only in the term with a single derivative and an overall factor $-1/2$. It therefore obeys the same commutation relations as $\tilde{\K}_{i,\lambda}$, i.e. $[\tilde{\D}, \hat{K}_\lambda] = i \hat{K}_\lambda$.

Considering the remaining terms, let us notice that we have:
\eas{
\tilde{F}^{\rm T}_{1, \phi_i} (k_i , 0) + \tilde{\Gamma}_{1, \phi_i}(k_i)
&
= i(- d_i - k_i \cdot \partial_{k_i} )= i\hat{D}_i
 \\[2mm]
\tilde{F}^{\rm T}_{2, \phi_i}(k_i, 0) + \tilde{\Gamma}_{2, \phi_i}(k_i) 
& =
- 2(D-2d_i) \partial_{k_i, \lambda} + \tilde{\K}_{\lambda, \phi_i}
= -2\hat{K}_{k_j,\lambda}
\label{FGK}
}
where for brevity we also defined $\hat{D}_i$, thus $[\hat{D}, \hat{K}_\lambda ] = \hat{K}_\lambda$.
From this we find that \Eq{doublesoftWI} reads:
 \eas{
& i2f_\xi^2 \partial_{k, \lambda} 
  \langle \txi(q) \txi(k) \tphi (k_1) \cdots
 \tphi (k_n)  \rangle_{\rm amp}\mid_{q,k \sim 0}
\\
&=
\Bigg [i (d_{\partial^2 \xi}+d_{x\partial^2 \xi} - 2D) 
\sum_{i=1}^n
 \hat{K}_{k_i, \lambda}
 +2
 \sum_{i=1}^n i \hat{D}_i \sum_{j\neq i} \hat{K}_{k_j, \lambda} 
 \\
&\quad
+ \sum_{i=1}^n i \hat{D}_i  \hat{K}_{k_i, \lambda} 
+ \sum_{i=1}^n i   \hat{K}_{k_i, \lambda} \hat{D}_i
\Bigg ]
\langle  \tphi(k_1) \cdots  \tphi(k_n) \rangle_{\rm amp} 
+\Ord(q,k)
 \\
&=
i\sum_{j=1}^n  \hat{K}_{k_j, \lambda} 
\Bigg [d_{\partial^2 \xi}+d_{x\partial^2 \xi}  - 2D + 1
 +2  \sum_{i=1}^n \hat{D}_i \Bigg]
\langle  \tphi(k_1) \cdots  \tphi(k_n) \rangle_{\rm amp} 
+\Ord(q,k)
 }
In going from the first equality to the second equality, we used the commutation relation between $\hat{D}$ and $\hat{K}_\lambda$.

Using that $d_{\partial^2\xi} = d + 2 = D-  d$ and $d_{x\partial^2\xi} = d + 1 = D-  d-1$, we arrive at:
\ea{
\begin{split}
 f_\xi^2 \partial_{k, \lambda} 
 & \langle \txi(q) \txi(k) \tphi (k_1) \cdots
 \tphi (k_n)  \rangle_{\rm amp}\mid_{q,k \sim 0}
 \\
 &
=
\sum_{j=1}^n  \hat{K}_{k_j, \lambda} 
\left (-d 
 +  \sum_{i=1}^n \hat{D}_i \right)
\langle  \tphi(k_1) \cdots  \tphi(k_n) \rangle_{\rm amp} 
+\Ord(q,k)
\end{split}
\label{dklambda}
}

It follows that by studying instead the Ward identity of $Q_1^\mu = \K^\mu$
 and $Q_2 = \D$, we equivalently find an expression reading
\eas{
 f_\xi^2 \partial_{q, \lambda} 
 & \langle \txi(q) \txi(k) \tphi (k_1) \cdots
 \tphi (k_n)  \rangle_{\rm amp}\mid_{q,k \sim 0}
 \\
 &
=
\sum_{j=1}^n  \hat{K}_{k_j, \lambda} 
\left (-d 
 +  \sum_{i=1}^n \hat{D}_i \right)
\langle  \tphi(k_1) \cdots  \tphi(k_n) \rangle_{\rm amp} 
+\Ord(q,k)
}
 which differs only from \Eq{dklambda} by the soft-momentum derivative 
on the left-hand side.
{Contracting} either expression with the respective soft momentum 
$k^\lambda$ and $q^\lambda$, it follows that these expressions provide
the $\Ord(q,k)$ terms in the Taylor series of the double-soft Ward identity.

Reducing these to relations among amplitudes, we use that only the dilatations give a contribution by acting on the momentum-conserving delta-function, thus yielding:
\ea{
& f_\xi^2 \partial_{k, \lambda} 
T_{n+2}(q,k,k_1, \ldots, \bar{k}_n)
=
\sum_{j=1}^n  \hat{K}_{k_j, \lambda} 
\left (D-d 
 +  \sum_{i=1}^n \hat{D}_i \right)T_n(k_1, \ldots, \bar{k}_n)
+\Ord(q,k)
\label{AmpWI}
}
and similarly for $\partial_{q, \lambda}$ acting on $T_{n+2}$.
By contracting these identities with $k^\lambda$ and $q^\lambda$ yields the soft expansion of $T_{n+2}$,
\ea{
T_{n+2}(q,k; k_i ) = T_{n+2}(0,0;k_i) + q\cdot \partial_q T_{n+2}(0,0;k_i) + k \cdot \partial_k T_{n+2}(0,0;k_i) + \cdots
}
which together with the result of the previous subsection explicitly reads:
\ea{
\begin{split}
f_\xi^2 &T_{n+2}(q,k,k_1, \ldots, \bar{k}_n)
=
\Bigg [
\left ( D- d + \sum_{i=1}^n \hat{D}_i \right )\left ( D + \sum_{i=1}^n \hat{D}_i \right )
 \\
&\qquad
+ (q^\lambda + k^\lambda) \sum_{i=1}^n \hat{K}_{k_i, \lambda} \left ( D - d + \sum_{i=1}^n \hat{D}_i \right )
\Bigg ]T_{n}( k_1, \ldots, \bar{k}_n) + \Ord(q^2,k^2,qk)
\end{split}
\label{fulldoublesoft}
}

 \subsection{Double-soft Ward identity of two special conformal currents: \\
 A no-go for higher-order soft factorization}
 We finally consider the double-soft Ward identity following two insertions of special 
conformal currents in the matrix element.
 We restrict again our attention to the part belonging only to the Taylor series 
of the soft expansion. Then since $[ \K_\mu, \K_\nu] = 0$ we may simply study 
 \Eq{masslessdoublesoftWI}.
 Using the identities in \Eq{DK} and \Eq{FGK} for the special conformal current, 
we can immediately write the double-soft Ward identity following from \Eq{masslessdoublesoftWI}:
 \ea{
 \begin{split}
& -4f_\xi^2 \partial_{q, \lambda} \partial_{k, \gamma}
  \langle \txi(q) \txi(k) \tphi (k_1) \cdots
 \tphi (k_n)  \rangle_{\rm amp}\mid_{q,k \sim 0}
\\[2mm]
&
=
- i \tilde{\K}_{\gamma, x_\nu \partial^2\xi}(q+k)
(i2f_\xi \partial_{q+k, \lambda} )
 \langle \txi(q+k)  \tphi (k_1) \cdots
 \tphi (k_n)  \rangle_{\rm amp}
\\
 &
 \quad - 
4 \sum_{i=1}^n
 \hat{K}_{\lambda, i}(k_i) 
\sum_{j =1}^n 
\hat{K}_{\gamma, i}(k_j)   
\langle  \tphi(k_1) \cdots  \tphi(k_n) \rangle_{\rm amp} 
+\Ord(q,k)
\end{split}
\label{doubleKK}
}
where $\tilde{\K}_{\gamma, x_\nu \partial^2\xi}(q+k)$ is defined 
in \Eq{Kmomentum}. 

This time we have run into a problem:
There is no single-soft Ward identity that relates the first term on the right-hand side to an expression in terms of the $n$-point correlation function. We have not been able to circumvent this problem, and it thus looks like a no-go theorem for obtaining soft factorization at the order $q_\mu k_\nu$. We furthermore note that we have no Ward identities that could potentially lead to soft factorization of terms with $q_\mu q_\nu$ and $k_\mu k_\nu$, which would be required to establish a full soft theorem at the order $q \, k$.
We note, however, that the second term does takes the form 
of a soft theorem, relating the $n+2$ point correlation function to the 
$n$-point function acted upon by two special conformal transformation.
 One may be able to express this for amplitudes as a relation between 
$n+2$-, $n+1$- and $n$-point function, but we do not attempt to do so 
here. 

 \section{Multi-soft dilatons}
\label{multisoft}

 In Sec.~\ref{singlesoft} we have derived the soft theorem for the emission of a single 
soft dilaton, through $\Ord(q)$ in the soft momentum, $q$, 
 while in Sec.~\ref{doublesoftgeneral} we have obtained a soft theorem for two soft 
dilatons 
through 
$\Ord(q_1^\mu q_2^\nu)$ with $q_1$ and $q_2$ the momenta of the two soft 
dilatons taken to be $q_1 \sim q_2 \ll k_i$, where $k_i$ is any of the hard momenta 
involved in the amplitude. In this section we will first show that the double-soft theorem 
is equivalent to what one would get by making two consecutive emissions of 
the soft dilatons, 
one  after the other, with $q_1 \ll q_2 \ll k_i$. 
 From this observation we can make the conjecture that the amplitude 
for the emission 
of any number of soft dilatons is fixed by the consecutive soft limit of single dilatons 
emitted one after the other, that is:
 \ea{
 \lim_{q_1, \ldots, q_m \to 0} A_{m+n}(q_1, \ldots, q_m; k_1, \ldots, k_n)
 = \lim_{q_1 \to 0}\lim_{q_{2} \to 0} \cdots \lim_{q_m \to 0} A_{m+n}(q_1, \ldots, q_m; k_1, \ldots, k_n)
 }
 where on the left-hand side it is assumed that all soft momenta scale 
simultaneously to zero, while on the right-hand side it is assumed 
that $q_m \ll q_{m-1} \ll \cdots \ll q_1 \ll k_i$.
 
 To see that this conjecture holds for the double-soft case, let us first summarize our previous results.
 The soft theorem for the emission of a single soft dilaton reads:
 \ea{
T_{n+1} (q, k_1, \ldots , \bar{k}_n) = &\frac{1}{f_\xi} \Bigg [ D 
+ \sum_{i=1}^n \hat{D}_i
+ q^\mu \sum_{i=1}^n {\hat{K}_{k_i, \mu}}
\Bigg ] 
T_n(k_1, \ldots, \bar{k}_n ) + \Ord(q^2)
}
The soft theorem for the simultaneous emission of two soft dilatons reads:
\eas{
f_\xi^2 &T_{n+2}(q_1,q_2,k_1, \ldots, \bar{k}_n)
=
\Bigg [
\left ( D- d + \sum_{i=1}^n \hat{D}_i \right )\left ( D + \sum_{i=1}^n 
\hat{D}_i \right )
 \\
&\qquad
+ (q_1^\lambda + q_2^\lambda) \sum_{i=1}^n {\hat{K}_{k_i, \lambda}} 
\left ( D - d + \sum_{i=1}^n \hat{D}_i \right )
\Bigg ]T_{n}( k_1, \ldots, \bar{k}_n) + \Ord(q_1^2,q_2^2, q_1 q_2)
}
where
\ea{
\hat{D}_i = - \left ( d_i  + k_i \cdot \partial_{k_i} \right ) 
\, , \qquad 
{\hat{K}_{k_i, \mu} =  \frac{1}{2} k_{i \mu} \partial_{k_i}^2
- (k_{i} \cdot \partial_{k_i} ) \partial_{k_i, \mu} 
- d_i \, \partial_{k_i, \mu} }
}
 
Now let us consider an $(n+2)$-point amplitude, which involves at least two 
dilatons, carying momenta $q_1$ and $q_2$. If we take the momentum 
$q_1$ to be soft compared to the other momenta, i.e. $q_1 \ll q_2, k_i$, then
the single soft theorem gives us:
\ea{
 & f_\xi T_{n+2}(q_1,q_2,k_1, \ldots, \bar{k}_n)
\label{pre}  \\
&=
   \left[ 
D + \sum_{i=1}^n \hat{D}_i - (d + q_2\cdot \partial_{q_2}) + q_1^\lambda 
\sum_{i=1}^n{ \hat{K}_{k_i, \lambda}} + 
q_1^\lambda \hat{K}_{q_2, \lambda} 
\right ]
  T_{n+1}(q_2, k_1, \ldots, \bar{k}_n) + \Ord(q_1^2)
 \nn
}
  If $q_2 \ll k_i$ in the above expression, the behavior of the $(n+1)$-point 
amplitude is also fixed through $\Ord(q_2^2)$, i.e.
   \ea{
T_{n+1} (q_2, k_1, \ldots , \bar{k}_n) = &\frac{1}{f_\xi} \Bigg [ D 
+ \sum_{i=1}^n \hat{D}_i
+ q_2^\mu \sum_{i=1}^n {\hat{K}_{k_i, \mu}}
\Bigg ] 
T_n(k_1, \ldots, \bar{k}_n ) + \Ord(q_2^2)
}
Inserting this expression in  Eq. (\ref{pre}) 
we find
  \eas{
   f_\xi^2 &T_{n+2}(q_1,q_2,k_1, \ldots, \bar{k}_n)
 \\
   &
  = 
  \left[ 
D + \sum_{i=1}^n \hat{D}_i - (d +q_2\cdot \partial_{q_2}) +
 q_1^\lambda \sum_{i=1}^n {\hat{K}_{k_i, \lambda}} + 
q_1^\lambda \hat{K}_{q_2, \lambda} 
\right ]
 \\
&\quad \times
\left [
D + \sum_{i=1}^n \hat{D}_i+ q_2^\lambda \sum_{i=1}^n {\hat{K}_{k_i, 
\lambda}}
\right ] T_{n}( k_1, \ldots, \bar{k}_n) + \Ord(q_1^2,q_2^2, q_1 q_2)
\\[5mm]
&=
\Bigg [
\left ( D- d + \sum_{i=1}^n \hat{D}_i \right )\left ( D + \sum_{i=1}^n 
\hat{D}_i \right )
+ \left ( D + \sum_{i=1}^n \hat{D}_i - d- 1 \right )   q_2^\lambda 
\sum_{i=1}^n {\hat{K}_{k_i, \lambda}}
 \\
&\quad
+    q_1^\lambda \sum_{i=1}^n {\hat{K}_{k_i, \lambda}} 
\left ( D + \sum_{i=1}^n \hat{D}_i \right )
- d\, q_1^\lambda \sum_{i=1}^n {\hat{K}_{k_i, \lambda}} 
\Bigg ]T_{n}( k_1, \ldots, \bar{k}_n) + \Ord(q_1^2,q_2^2, q_1 q_2)
} 
After the second equaltiy, the first three terms are just an organized expansion of the multiplication, where 
the form of $\hat{D}_i$ and {$\hat{K}_{k_i,\lambda}$} 
is unimportant and one only needs to use in the second term the identity
$q_2\cdot \partial_{q_2} q_2^\lambda = q_2^\lambda$. 
The last term is obtained by using $q_1^\lambda \hat{K}_{q_2,\lambda} 
\, q_2^\rho = q_1^\rho(- d )$.  The term of  order $q_1^\lambda q_2^\rho$ has 
been neglected.

Using the commutation relation $[\hat{D}_i , {\hat{K}_{k_i, \lambda} }] = {
\hat{K}_{k_i, \lambda}}$, the expression reduces to:
\eas{
f_\xi^2 &T_{n+2}(q_1,q_2,k_1, \ldots, \bar{k}_n)
=
\Bigg [
\left ( D- d + \sum_{i=1}^n \hat{D}_i \right )\left ( D + \sum_{i=1}^n \hat{D}_i \right )
 \\
&\qquad
+ (q_1^\lambda + q_2^\lambda) \sum_{i=1}^n {\hat{K}_{k_i, \lambda}} 
\left ( D - d + \sum_{i=1}^n \hat{D}_i \right )
\Bigg ]T_{n}( k_1, \ldots, \bar{k}_n) + \Ord(q_1^2,q_2^2)
}
thus exactly reproducing the double-soft theorem \Eq{fulldoublesoft} derived from current algebra.
Based on this result, we conjecture that multi-soft dilaton amplitudes are fixed by 
the consecutive soft limit of single dilatons emitted one after the other, as just 
detailed for the consecutive double-soft emission.

 \section{Examples of dilaton amplitudes}
 \label{dilatonamplitude}

 \subsection{The simplest $D$-dimensional conformally broken field theory}
 
 We consider amplitudes of the simplest $D$-dimensional conformal model presented in Sec.~\ref{hidden}, and specifically given by \Eq{simplecft}.
In the spontaneously broken phase, the Lagrangian is expanded around a nonzero vacuum expectation value for the conformal compensator field $\bar{\xi} = f_\xi/d + \xi$, where $\xi$ is the dilaton field, and $f_\xi = d v^d$,
\ea{
\mathcal{L} = - \frac{1}{2}(\partial_\mu \chi)^2 - \frac{1}{2} (\partial_\mu \xi)^2 
- \frac{1}{2} m^2 \chi^2 
- \frac{m^2}{f_\xi} \chi^2 \xi
- \frac{c_2}{2} \frac{m^2}{f_\xi^2}  \chi^2 \xi^2
- \frac{c_3}{3!} \frac{m^2}{f_\xi^3}  \chi^2 \xi^3
- \frac{c_4}{4!} \frac{m^2}{f_\xi^4}  \chi^2 \xi^4
+ \ldots \label{ac}
}
where the mass is related to the dimensionless coupling constant and vev in the following manner:
\ea{
m^2 =  v^2 \lambda^{2/d}
}
and the first few coefficients read:
\ea{
c_2 &= \frac{6-D}{2 } \, , \quad
c_3 
= \frac{(6-D)(4-D)}{2} \, , \quad
c_4 
=  \frac{(6-D)(4-D)(10-3D)}{4} 
}
having used that $d = [\xi] = (D-2)/2$.

We have expanded the Lagrangian up to the
sixth order in the fields, since we would now like to 
compute the three-, four-, five- and six-point amplitudes involving two 
massive external 
states 
$\chi$, and one, two, three and four dilatons, respectively. The three point amplitude is 
given by
the only three point vertex, reading:
\ea{
T_3^{2\chi, \xi} = - \frac{ 2 m^2}{f_\xi} = - \frac{4}{D-2} \frac{ m^2}{ v^d} \, .
}
There are no derivative couplings in the Lagrangian. Thus momenta enter amplitudes only from internal propagators. For amplitudes with two massive external states, only massive internal propagators enter.
It is useful to define the variables
\ea{
s_{i_1, i_2, \ldots i_n}  = (k_{i_1} + \cdots + k_{i_n} )^2 + m^2
}
where the indices enumerate the external states. We will take the two massive state to be states 1 and 2, thus entering amplitudes with momenta $k_1$ and $k_2$, while states $3, \ldots, n$ are taken to be dilatons entering amplitudes with momenta $k_3, \ldots, k_n$.

The four-point amplitude then reads:
\ea{
T_4^{2\chi, 2\xi} = - \frac{2 m^2}{f_\xi^2} \left (
 c_2 - \frac{2m^2}{s_{13}} - \frac{2m^2}{s_{23}} \right )
\label{T4}
}
which has contributions both from the four-point vertex and two three-point amplitudes attached by an internal massive propagator. Momentum conservation is implicit in this expression, e.g. $s_{13} = s_{24}$.

The five-point amplitude reads:
\ea{
\begin{split}
T^{(2\chi, 3\xi)}_5 =& - c_3 \frac{2m^2}{f_\xi^3} + c_2 \frac{\left( 2 m^2 \right)^2}{f_\xi^3}
  \left[ 
  \frac{1}{s_{13}} + \frac{1}{s_{23}} 
  +
\frac{1}{s_{14}} +
\frac{1}{s_{24}}+  \frac{1}{s_{15}} + \frac{1}{s_{25}} 
 \right]
\\
& - \left( \frac{2m^2 }{f_\xi} \right)^3 \left[ 
 \frac{1}{s_{14}s_{23}} 
 +  \frac{1}{s_{24}s_{13}}  
+  \frac{1}{s_{24}s_{15}}  
+  \frac{1}{s_{14}s_{25}}  +
 \frac{1}{s_{25}s_{13}}  
+  \frac{1}{s_{15}s_{23}}  
 \right]
 \end{split}
\label{T5}
}
Finally, the six-point amplitude reads:
\eas{
T^{(2\chi, 4\xi)}_6=& -\frac{2c_4}{f_\xi^4}+c_3\frac{(2m^2)^2}{f_\xi^4}\sum_{i=3}^6\left[\frac{1}{s_{1i}}+\frac{1}{s_{2i}}\right]+c_2^2\frac{(2 m^2)^2 }{f_\xi^4} \sum_{i=4}^6 \left[\frac{1}{s_{13i}}+\frac{1}{s_{23i}}\right] 
\\
&-c_2\frac{(2m^2)^3}{2f_\xi^4} \sum_{i=3}^6\left[\frac{1}{s_{1i}}\sum_{j\neq 1,2,i} \left(\frac{1}{s_{2j}}+\frac{2}{s_{1ij}}\right)+ \frac{1}{s_{2i}}\sum_{j\neq 1,2,i} \left(\frac{1}{s_{1j}}+\frac{2}{s_{2ij}}\right)\right]
\\
&+\frac{(2m^2)^4}{f_\xi^4} \sum_{i=3}^6\frac{1}{s_{1i}  } \sum_{j\neq 1,2,i}\frac{1}{s_{2j}}\sum_{k\neq 1,2,i,j}\frac{1}{s_{1ik}}\label{6point}
}
The soft theorems provided in this work can now all be explicitly checked.
Some details should be noted. First, one must fix an overall momentum variable by momentum conservation. Since we are interested in the expansion of the soft momenta, we do not impose momentum conservation on these variables, but instead impose it on one of the hard dilaton momenta. For instance, taking the momenta $k_5$ and $k_6$ to be soft for relating the 6-, 5- and 4-point amplitudes, a consistent choice is to take
\ea{
k_4 \to \bar{k}_4 
}
where $\bar{k}_4$ is replaced by minus the sum of all other momenta of the 4-, 5- and 6-point amplitudes. This is already explicit in \Eq{T4} for $T_4$, and is trivially imposed on the 5- and 6-point amplitudes, e.g. $s_{14} \to s_{235}$ in $T_5$ or $s_{14} \to s_{2356}$ in $T_6$.

The next important step one must make to check our expressions, is to subtract from the amplitudes all terms that belong to the Laurent series in the soft expansion, as defined in \Eq{Flaurent}.
For instance, considering the single soft limit of $T_5$ when $k_5 \ll k_i$ for $i = 1, \ldots,4$, the part of $T_5$ that gives the Taylor series in $k_5$ is:
\ea{
&T^{(2\chi, 3\xi)}_{5, \rm Taylor} = - c_3 \frac{2m^2}{f_\xi^3} + c_2 \frac{\left( 2 m^2 \right)^2}{f_\xi^3}
  \left[ 
  \frac{1}{s_{13}} 
  +
\frac{1}{s_{135}}
 \right]
 - \left( \frac{2m^2 }{f_\xi} \right)^3 \left[ 
  \frac{1}{s_{13} s_{135}}  
 \right] + (s_{1 \ldots} \leftrightarrow s_{2\ldots})
\label{T5taylor}
}
It is now obvious that at leading order in $k_5$, this expression reads:
\ea{
&T^{(2\chi, 3\xi)}_{5, \rm Taylor} = - \frac{2m^2}{f_\xi^3} 
\left [c_3
-  2 c_2 
  \frac{2 m^2}{s_{13}}
 + 
  \frac{(2m^2)^2}{s_{13}^2}  +  (s_{1 \ldots} \leftrightarrow s_{2\ldots})
  \right ] + \Ord(k_5)
\label{T5leadingtaylor}
}
It is a straightforward exercise from here to check that:
\ea{
\frac{1}{f_\xi} \left[ D- 4 d - \sum_{i=1}^3 k_i \cdot \partial_{k_i} \right ] T_4^{(2\chi, 2\xi)} (k_1, k_2, k_3,
\bar{k}_4
 ) = T^{(2\chi, 3\xi)}_{5, \rm Taylor} + \Ord(k_5)
} 
where $d=(D-2)/2$, in agreement with \Eq{leadingsoft}. We remark that this expression also takes into account the massive terms in \Eq{T5leadingtaylor}. The full expression for $T_5$ also has contributions at $\Ord(k_5^0)$ from expanding terms such as
\ea{ 
\frac{1}{s_{15}s_{135}} = \frac{1}{2(k_1 \cdot k_5) s_{13}} \left ( 1 + \frac{2 (k_1 + k_3)\cdot k_5}{s_{13}}\right ) + \Ord(k_5)
}
however, these terms belong to the Laurent series of the soft expansion, and thus not part of \Eq{leadingsoft}.

The check of the single soft theorem is now extended to the  subleading order of the five point amplitude.  The $O(k_5)$ terms of the five point amplitude read:
\ea{
T_{5\,\rm Taylor}^{(2\chi,\,3\xi)}\Big|_{O(k_5)}=-2\frac{(2m^2)^2}{f_\xi^3}
 \frac{k_5\cdot (k_1+k_3)}{s_{13}^2}
\Bigg[\frac{(6-D)}{2}-\frac{2m^2}{s_{13}}\Bigg]+(1\leftrightarrow 2)
}
and it is straightforward to verify that it satisfies the identity:
\ea{
& \frac{k_5^\mu}{f_\xi}\sum_{i=1}^3\Bigg[ \frac{1}{2} k_{i\mu} \frac{\partial^2}{\partial k_{i\nu}\partial k_i^\nu}
-k_i^\nu\frac{\partial}{\partial k_i^\mu}\frac{\partial}{\partial k_i^\nu} -
d \frac{\partial}{\partial k_i^\mu}
\bigg] T_4^{2\chi;2\xi}(k_1,k_2,k_3, 
\bar{k}_n
)= T_{5\,\rm Taylor}^{(2\chi,\,3\xi)}\Big|_{O(k_5)}
}
in agreement with the single soft theorem in \Eq{subleadingsoft}, {
as originally proposed in Ref.~\cite{DiVecchia:2015jaq}}.

The single-soft dilaton relations between $T_6$ and $T_5$ can be checked in a similar fashion.

The double-soft relations between the 5- and 3-point amplitudes can be easily verified.
Choosing the soft momenta to be $k_4,\,k_5\ll k_i$, $i=1,2,3$,  and using momentum 
conservation to   replace $k_3$ with the other momenta, we first notice that only 
the first term in Eq. (\ref{T5}) is regular in the double-soft limit; i.e. all other terms (which carry the momentum dependence) belong to the Laurent series of the soft-expansion and should not be considered. It is then easy to see that
\ea{
T_{5;\rm Taylor}^{(2\chi,3\xi)}=\frac{(6-D)(4-D)}{2\,f_\xi^2} T_3^{(2\chi,\xi)} =
\frac{1}{f_\xi^2} (D-4d) (D-3d) T_3^{(2\chi,\xi)}
}
where $d=(D-2)/2$ is the scaling dimension of all the fields. 
Since $T_3^{(2\chi,\xi)}$ is momentum independent, this expression is
exactly the prediction of the double-soft theorems, both the one coming 
from the Ward identity of two dilatation currents, but also (trivially) the one 
coming from a dilatation current and a special conformal transformation
current, since $\partial_{4,5}^\mu T_{5;\rm Taylor}^{(2\chi,3\xi)} = 
\hat{K}_i^\mu T_3^{(2\chi,\xi)}  = 0$. This example also shows, how 
reversibly one can predict coefficients of effective actions from the 
soft theorems, here a 5-point interaction coefficient from knowledge 
of the three-point interaction.

Before making the similar checks on the much less trivial case of 6- and 4-point amplitudes,
let us note that the rest of $T_5$, which are on-shell singular for $k_4=k_5=0$, read
\eas{
&T_{5;\rm Laurent}^{(2\chi,3\xi)} =\left(
\frac{1}{s_{15}}+\frac{1}{s_{14}}+\frac{1}{s_{245}} +\frac{1}{s_{24}}+\frac{1}{s_{25}} +\frac{1}{s_{245}}\right)
T_{3}^{(2\chi,\xi)}V_{4}^{(2\chi,2\xi)}
\\
&+(T_{3}^{(2\chi,\xi)})^3\left[\frac{1}{s_{14}}\left( \frac{1}{s_{23}}+\frac{1}{s_{25}}\right)+\frac{1}{s_{15}}\left( \frac{1}{s_{24}}+\frac{1}{s_{23}}\right)\frac{1}{s_{245}}\left( \frac{1}{s_{24}}+\frac{1}{s_{25}}\right)\right]
}
where we identified the 4-point vertex $V_{4}^{(2\chi,2\xi)}= -\frac{2m^2}{f_\xi^2}c_2$. 
In this form, it is easy to see that all terms belonging to the Laurent series of the double-soft expansion are simply coming from processes where two soft dilatons are directly emitted from the hard external legs in different ways. This observation applies generally to all tree-amplitudes and trivializes thus the Laurent part of the soft-expansion.

We now consider the double-soft expansion of $T_6$  in terms of the soft momenta 
$k_5$ and $k_6$ through order $\Ord(k_5, k_6)$. The softness of the two momenta 
are taken to be equal, and we should thus consider the  Taylor expansion of $T_6$ 
around $(k_5, k_6) = (0,0)$. As prescribed we need to replace $k_4 \to \bar{k}_4$ and 
remove terms that belong to the Laurent series. From \Eq{6point} we then find:
\ea{
&T_{6, \rm Taylor}^{(2\chi, 4\xi)}=
- c_4 \frac{2m^2}{f_\xi^4} + 
\left(\frac{2m^2}{f_\xi^2}\right)^2
\Bigg[\frac{c_3}{s_{13}} +\frac{c_3}{s_{1356}}+\frac{c_2^2}{s_{135}}+\frac{c_2^2}{s_{136}}+\frac{(2m^2)^2}{s_{13}s_{1356}}\Big(\frac{ 1}{s_{135}}+\frac{ 1}{s_{136}}\Big)  \nonumber\\
&
 -  c_2\Big(\frac{2m^2}{s_{13} s_{1356}} +\frac{2m^2}{s_{13}s_{135}}+\frac{2m^2}{s_{13}s_{136}} +\frac{2m^2}{s_{136}s_{1356}}+\frac{2m^2}{s_{135} s_{1356}}\Big)+(s_{1\dots} \leftrightarrow s_{2\dots})\Bigg]
}
From here it is straightforward to show that the Taylor-expansion of this expression through 
first order around $(k_5, k_6) = (0,0)$ exactly match the double-soft theorem in 
\Eq{fulldoublesoft}, by using  the four-point amplitude in \Eq{T4}.

For completeness, we note again that the on-shell singular terms for $(k_5,\,k_6)=0$; i.e. those belonging to the Laurent expansion of the amplitude,  can be compactly  written as:
\ea{
\begin{split}
&T_{6, \rm Laurent}^{(2\chi, 4\xi)}= \left[T_3^{(2\chi,\xi)} \frac{1}{s_{15}} T_5^{(2\chi;3xi)}(k_1+k_5, k_2,k_3,k_4,k_6)+(5\leftrightarrow 6)+(1\leftrightarrow 2)\right]
\\
&\quad +\left[ \frac{ V_4^{(2\chi,2\xi)}}{s_{256}}T_4^{(2\chi,2\xi)}(s_{13}, s_{14})
-\frac{T_3^{(2\chi,\xi)}}{s_{15}}T_4^{(2\chi,2\xi)}(s_{135},s_{263})\frac{T_3^{(2\chi,\xi)}}{s_{26}}+(1\leftrightarrow 2)\right]
\end{split}
\label{5sing}
}
The terms in the first line corresponds to the cases where a soft dilaton is 
directly emitted from one of the hard, massive, external states, through the 
3-point interaction vertex, which is equivalent to the amplitude $T_3$. 
The similar type of process where two soft dilatons are emitted from the hard, massive 
legs are given in the second line, involving two factors of $T_3$, while finally
the case corresponding to the process where two soft dilatons are emitted simultaneously 
and from the same point from a hard, massive external state is also present and involves the 4-point 
interaction vertex, $V_4 = - c_2 (2m^2)/f_\xi^2$.

 \subsection{$\mathcal{N}=4$ super Yang-Mills theory on the Coulomb branch}
 \label{N=4}
The $\mathcal{N}=4$ supersymmetric Yang-Mills (SYM) theory is a (super)conformal field theory, where the gauge coupling stays nonperturbatively unrenormalized. Its action in component fields of the supermultiplet reads:
\ea{
&S = \int d^4 x {\rm Tr} \left(- \frac{1}{4} G_{\mu \nu} G^{\mu \nu}  - \frac{1}{2}
(D_\mu \phi_i)^2 + \frac{i}{2} {\bar{\psi}^r} \gamma^\mu D_\mu \psi_r
+ \frac{g}{2} {\bar{\psi}^r}
 \Gamma_{rr'}^i [ \phi_i , \psi^{r'}] + 
\frac{g^2}{4} 
\left([ \phi_i , \phi_j ]^2  \right) \right)\nonumber \\
\label{SN=4}
}
where $r,r' = 1, \ldots, 4$, $i,j =1 \dots 6$, $D_\mu = \partial_\mu - i g [A_\mu,  \cdot]$ and 
$\Gamma_i$ are Euclidean six-dimensional Dirac matrices satifying the 
anti-commutation relations $\{\Gamma_i , \Gamma_j\} = \pm 2 \delta_{ij}$.  
All fields are in the adjoint representation of the gauge group.
The theory possesses an $SU(4)$ global $R$-symmetry, under 
which the fermions transform in the fundamental, $\bf{4}$, representation 
and the scalars transform in the antisymmetric rank two, $\bf{6}$,
representation. 
The potential is given by
\ea{
{\rm Tr} ([\phi_i, \phi_j][\phi_i, \phi_j] )= -f^{abe}~f^{cde} ~
\phi_i^a \phi_j^b \phi_i^c \phi_j^d 
}
where we have used $\phi_i = \phi_i^a T^a$, $[T^a, T^b] = i f^{abc} T^c$,
 and ${\rm Tr} (T^aT^b) = \delta^{ab}$.
If $a=b$ or $c=d$ then this expression vanishes, due to antisymmetry of the structure constant $f^{abe}$. 
This is independent of the value of $\phi_i$ and thus there is an $O(6)$ 
symmetry of this minimum.
Any vev acquired by one of the scalars, breaks spontaneously the 
 conformal symmetry and the $SU(4)$ global R-symmetry, isomorphic to $SO(6)$ (under which the scalars 
transform as vectors),
is broken to $SU(4) \to Sp(4)$ (or equivalently $SO(6) \to SO(5)$). This is 
the so-called Coulomb branch of the theory. There will be 5 
Nambu-Goldstone (NG) bosons belonging to the breaking of the global group, 
and one additional NG boson belonging to the breaking of conformal symmetry, 
i.e. \emph{the dilaton}. 

The gauge symmetry is also broken, but the additional gauge degrees of freedom of the scalars will be eaten up by the corresponding gauge bosons. 
To be specific, consider the $SU(N+1)$ gauge theory. The Coulomb branch  
induce $SU(N+1) \to SU(N)\times U(1)$. At low energies where massive states 
decouple, the $SU(N)$ and $U(1)$ sectors are two separate SYM theories, where 
the 6 NG bosons form the 6 massless scalars of the $U(1)$ supermultiplet.

The Ward identities and soft theorems presented in this work, can be checked to be satisfied by explicit computation of amplitudes in the weakly coupled regime of the above action on the Coulomb branch. For the single-soft dilaton, the check has been performed in Ref.~\cite{Bianchi:2016viy} through one loop.
Here we will instead consider the strongly coupled regime of the theory on 
the Coulomb branch by utilizing its gravity dual, for instance 
described in Section 6 of Ref.~\cite{Elvang:2012st}.

The gravity dual of the Coulomb branch is modeled by a D3-probe brane in the gravitational background of $N$ D3-branes. In the large $N$ limit backreaction on the background can be neglected. 
The dynamics of the D3 brane is governed by the Dirac-Born-Infeld (DBI) action on AdS${}_5\times S^5$, which including the Wess-Zumino term for the zero-force condition (the pullback of the 5-form flux), is given by:
\ea{
&S = - \frac{1}{\kappa^2}  \int d^4 x \frac{r^4}{L^4} \left(
 \sqrt{ - \det \left( 
\eta_{\mu \nu} + \frac{L^4}{r^4}
\frac{\partial x^i}{\partial x^\mu} 
 \frac{\partial x^i}{\partial x^\nu}  
+ \kappa \frac{ L^2}{r^2}  F_{\mu \nu}
\right) }  -1 \right)
\label{SDBI}
}
where $\kappa = (2 \pi)^{3/2} \alpha' \sqrt{ g_s} $,  $L$ is the AdS${}_5$ radius, $r^2 = 
\sum_{i=1}^6 x_i^2$ is the $S^5$ radius, $\eta_{\mu \nu}$ is the metric on the D3-brane with indices $\mu, \nu = 0, \ldots, 3$  and $x_i$ are the bulk coordinates with $i = 4, \ldots 9$.
The scalar field dynamics on the D3-brane is given by correctly normalizing the bosonic coordinates
\ea{
x_i = \kappa \phi_i \, , \quad \phi^2 = \sum_{i=1}^6 \phi_i^2
}
leading to 
\ea{
S=  - \frac{1}{\lambda^2}  \int d^4 x  \,\, \phi^4 \left(
 \sqrt{ - \det \left( 
\eta_{\mu \nu} + \frac{\lambda^2}{\phi^4}
\frac{\partial \phi^i}{\partial x^\mu} 
 \frac{\partial \phi^i}{\partial x^\nu}  
+ \frac{ \lambda}{  \phi^2}  F_{\mu \nu}
\right) }  - 1 \right)
\label{SDBI4}
}
where
\ea{
\lambda \equiv \frac{L^2 }{\kappa}
}
We note that $\lambda$ is a dimensionless constant. 
Using the dictionary relating the $AdS_5$ radius with  
the gauge coupling constant, one finds that $\lambda$ is fixed by the 
{$SU(N)$}
gauge group of the $\mathcal{N} = 4$ SYM dual as follows
\ea{
\frac{L^4}{{\alpha'}^2} = 4\pi N g_s \, \quad \Rightarrow \ \lambda = \frac{\sqrt{2N}}{2 \pi} 
\label{lam}
}
The previous action is
conformally invariant and is well-defined locally only if one of the scalar fields gets a 
non-vanishing  vacuum expectation value that breaks spontaneously
the conformal symmetry. Such a field with a non-vanishing vev will be the
dilaton, while the other five scalar fields should describe the NG
bosons corresponding to the breaking of the R-symmetry group $SO(6) \to SO(5)$. 

In this setup, the Ward identities and soft theorems proposed in this work should be satisfied.
We will here describe the check on the relations between the 4-, 5-, and 6-point dilaton tree amplitudes.
(We note that as an effective field theory, only tree amplitudes of this theory are supposed to describe the $\mathcal{N}=4$ SYM theory in the strongly coupled regime.)
It is to this end only necessary to consider the part of the Lagrangian involving the dilaton 
field up to six-point interactions. We choose to take the following Coulomb branch:
 \ea{
 \phi_i = v \delta_{i6} + {{\tilde{\phi}}_i} \, , \quad {{\tilde{\phi}}_6} \equiv \xi
\label{vev}
}
Then expanding the action, we find the following interaction Lagrangian for the dilaton 
\ea{
\mathcal{L}_{4,5,6}^{\xi} = \frac{\lambda^2}{8v^4} \left[1 - \frac{4 \xi}{v} + 
10 \frac{\xi^2}{v^2}   \right]    \left(\partial_{\mu} \xi \partial^\mu \xi   
\right)^2  -   \frac{\lambda^4}{16 v^8 }\left( \partial_\mu \xi 
\partial^\mu \xi  \right)^3 
\label{L6dila}
}
describing dilaton self-interactions up to six-points. 

It is straightforward to compute the four-point amplitude simply given by the contact interaction above. It reads:
\ea{
A_4 = \frac{\lambda^2}{4 v^4} [ s_{12} s_{34} + s_{13} s_{24}+s_{14} s_{23} ]  = 
\frac{ 4 \Delta a}{v^4 } [ s^2 + t^2 + u^2]
}
where $s_{ij} = (k_i + k_j)^2$ and in the second equality we identified the so-called $\Delta a = 16 \lambda^2 = N^2 /(8 \pi)^2$ parameter of the works on the dilaton effective action and a-theorem~\cite{Schwimmer:2010za, Komargodski:2011vj, Elvang:2012st}, as well as the Mandelstam variables, $s = - s_{12}$, $t = - s_{13}$, $u = - s_{23}$, after imposing momentum conservation. The five-point amplitude is also straightforwardly computed from the contact interaction only, and is simply related to the four-point amplitude as follows:
\ea{
\begin{split}
A_5(1,2,3,4,5) = - \frac{4}{v} \Big [ &A_4(1,2,3,4)
+A_4(1,2,3,5) + A_4(1,2,4,5)
 \\
& + A_4(1,3,4,5)+A_4(2,3,4,5) \Big]
\end{split}
\label{7.30}
}
Finally, we provide the expression for the six-point amplitude. The computation is more involved, since there are contributions from three different interactions, where two involve the two different six-point contact interactions and one involve two four-point interactions where one dilaton is exchanged between them, thus containing an on-shell pole.
Accordingly, we divide the amplitude in three partial expressions in the following way:
\ea{
A_6 = \lambda^2 A_6^{\partial^4} +\lambda^4 \left (A_6^{\partial^6} + A_6^{\rm pole} \right )
}
where we defined the partial amplitudes without the coupling constant, to make explicit the different powers it enters with. It follows that since $A_5$ and $A_4$ only contain terms with $\lambda^2$ couplings, only the first partial amplitude is related to the lower-point amplitudes through the soft theorems. 
The soft theorems thus immediately predict that the two other partial amplitudes should either cancel or vanish in the soft limits.

The pole terms are straightforwardly given in terms of the four-point amplitude as follows:
\ea{
\lambda^4 A_6^{\rm pole}
= \sum_{\text{ineq. perm.}} \frac{A_4(1, 2, 3, - [123] ) A_4([123], 4,5,6)}{s_{123}}
}
where the entry $[123]$ indicates that the momentum variable is equal 
to \mbox{$(k_1 + k_2 +k_3)$}, which due to momentum conservation  is the momentum 
exchanged between the two vertices, explaining also the denominator (propagator).
The sum is over the 10 inequivalent ways of choosing three out of the 6 momenta modulo the complement.
The order is unimportant, since $A_4$ is totally symmetric in the four momenta.
We can denote the 10 terms by their pole structure, given by:
\ea{
\{ s_{123}, \ s_{124}, \ s_{125}, \ s_{126}, \ s_{134}, \ s_{135}, \ s_{136}, \ s_{145}, \ s_{146}, \ s_{156} \}
}
The partial amplitude $A_6^{\partial^4}$ can also be given in terms of $A_4$ in the following way:
\ea{
\lambda^2 A_6^{\partial^4}=
\frac{20}{v^2} \sum_{\rm cycl. perm}^{1,\ldots 6} \left [
A_4(1,2,3,4)
+
A_4(1,2,3,5) +
\frac{1}{2} A_4(1,2,4,5)  \right ]\label{7.34}
}
where the sum is over cyclic permutations of the indices $1,2,3,4,5,6$ generating six terms from each of the above three terms. The factor $1/2$ on the last term is due to the extra symmetry of that term, and thus takes care of overcounting of the sum.

Finally, the expression for the partial amplitude $A_6^{\partial^6}$ reads:
\ea{
\lambda^4 A_6^{\partial^6}=
 \frac{3 \lambda^4}{8 v^8} \sum_{\rm cycl. perm}^{1,\ldots 6}\Bigg [
\frac{s_{14}s_{25}s_{36}}{6}
+ \frac{s_{12}s_{34}s_{56}}{3} 
+ \frac{s_{14}s_{23}s_{56}}{2}
+\frac{s_{15}s_{24}s_{36}}{2}
+ s_{13}s_{24}s_{56}
\Bigg]
}
where the denominators of the terms in the bracket indicate the permutation symmetry of the terms to avoid overcounting, e.g. the first term reproduces itself by any of the 6 cyclic permutations.

We now study the single-soft and double-soft dilaton properties of these amplitudes.
To study the relations between the 5- and 4-point amplitudes,
 we first fix momentum conservation and replace overall the momentum $k_4$ with minus the sum of the other momenta. 
It then becomes a straightforward exercise to check the following relations:
\ea{
\begin{split}
\lim_{k_5 \to 0 } A_5 (1,2,3,\bar{4}, 5) &= \frac{1}{v} \left [ 4- \sum_{i=1}^4 (d_i + k_i \cdot \partial_{k_i} )\right ]
A_4(1,2,3,\bar{4}) 
\\
&= 
-\frac{1}{v} \sum_{i=1}^3 k_i \cdot \partial_{k_i}
A_4(s,t,u) 
 = -\frac{4}{v}A_4(s,t,u)
 \end{split}
 \label{leading54}
  \\[2mm]
 \begin{split}
 \lim_{k_5 \to 0 } \partial_5^\mu A_5 (1,2,3,\bar{4}, 5) &= \frac{1}{v} 
  \sum_{i=1}^4 \hat{K}_{k_i}^\mu
A_4(1,2,3,\bar{4}) 
= 
\frac{1}{v} \sum_{i=1}^3  \hat{K}_{k_i}^\mu
A_4(s,t,u)  
 \\
& = -\frac{2\lambda^2}{v^5} [ s_{23}\, k_1^\mu + s_{13}\, k_2^\mu + s_{12} \, k_3^\mu ]
\end{split}
\label{subleading54}
}

To study the similar relations between the 6- and 5-point amplitudes we take $k_6$ to be soft. It is readily seen that $A_6^{\partial^6}$ and $A_6^{\rm pole}$ do not contribute to the soft limit $k_6 \to 0$ of $A_6$, since they contain in each term the soft momentum $k_6$. 
This is consistent with the observation made before that these two contributions should either vanish or cancel in the soft limits. 
The leading order single-soft relation between $A_6$ and $A_5$ is easiest to check by not imposing momentum conservation. It is then easy to confirm that:
\eas{
&\lim_{k_6 \to 0 } A_6 (1,2,3,{4}, {5}, 6) = 
\lambda^2 \lim_{k_6 \to 0 } A_6^{\partial^4} 
=
\frac{1}{v} \left [ -1- \textstyle{\sum_{i=1}^5} k_i \cdot \partial_{k_i} \right ]
A_5(1,2,3,{4},{5})
 \\
&= 
\frac{20}{v^2}
\Big [ A_4(1,2,3,4)
+A_4(1,2,3,5) + A_4(1,2,4,5)
 + A_4(1,3,4,5)+A_4(2,3,4,5) \Big]
}
where the second equality readily follows from $\sum_{i=1}^5 k_i \cdot \partial_{k_i} A_5 = 4 A_5$.
This works without the need to impose momentum conservation, because every term is linear in each momentum.

The subleading single-soft relation between $A_6$ and $A_5$ implies the two relations:
\sea{
\lim_{k_6 \to 0} \lambda^2 \partial_6^\mu A_6^{\partial^4}  &= \frac{1}{v} \sum_{i=1}^5  \hat{K}_{k_i}^\mu A_5 \\
\lim_{k_6 \to 0}  \lambda^4 \partial_6^\mu (A_6^{\partial^6}  + A_6^{\rm pole} ) &= 0
\label{cancellation}
}
As explained before, the reason for having two relations is clear by noting that $A_5$ only involves 
terms with $\lambda^2$ couplings.
The first relation can be seen as a constraint on the four-derivative interaction 
term from the five-point interaction. The second relation can be seen as 
a constraint on the six-derivative interaction term from the four-point 
interaction, because the pole terms are composed of two four-point vertices. 
The latter relation,  which involves cancellation of poles}, is nontrivially
satisfied, and we have shown this in detail in 
the Appendix. We will here show in some detail the validity of the first relation.
By expanding \Eq{7.34} at the first order in the soft momentum $k_6$, we get after some rewriting: 	
\eas{
	\lambda^2 A_6^{\partial^4}&(1,2,3,\bar{4},5,6)\big|_{O(k_6)}=-\frac{20}{v^2}\Bigg[ A_4(1,2,3,6)+A_4(1,2,5,6)+A_4(1,3,5,6)\\
	&+A_4(2,3,5,6)+\frac{\lambda^2}{v^4} (k_1 +k_2+k_3+k_5)^2(k_1+k_2+k_3+k_5)k_6\Bigg]\\
	=&-\frac{20}{v^2}\Big[ A_4(1,2,3,6)+A_4(1,2,5,6)+A_4(1,3,5,6)+A_4(2,3,5,6)\Big]\label{A.17}
}
	where  the  second equality follows from the identity 
$(k_1 +k_2+k_3+k_5)^2(k_1+k_2+k_3+k_5)k_6=
-2(k_4k_6)(k_4k_6)=0+O(k_6^2)$.

On the other hand, the action of the subleading soft operator on the five point amplitude can be seen to give:
\eas{
	\frac{k_6^\mu}{v}\sum_{i\neq 4}^{5}&\hat{K}_{k_i,\mu} \, A_5(1,2,3,\bar{4},5,6)
=-\frac{16\lambda^2}{v^6}\Big[k_6(k_1+k_2+k_3+k_5)(k_4k_6)\Big]
	\\
	&
	-	\frac{20}{v^2}\Big[A_4(1,2,3,6)+A_4(1,2,5,6)+A_4(1,3,5,6)
	+A_4(2,3,5,6)\Big]\\
	=&-\frac{20}{v^2}\Big[A_4(1,2,3,6)+A_4(1,2,5,6)+A_4(1,3,5,6)+A_4(2,3,5,6)\Big]+O(k_6^2)\label{A.18}
}
We observe that, as predicted, \Eq{A.17} and \Eq{A.18} are identical.

Moving on to the double-soft theorems, we here check the newly obtained relations between the 
6- and 4-point amplitudes.
We fix $k_4$ by momentum conservation in both amplitudes, and take $k_5$ and 
$k_6$ to be soft momenta. We note that $A_6^{\partial^6}$ 
and $A_6^{\rm pole}$ (except for Laurent terms) do not contribute to the soft limit $k_5, k_6 \to 
0$ of $A_6$ nor $\partial_{5,6}^\mu A_6$, since they contain in each term both momenta $k_5$ and $k_6$. 
The Laurent terms in $A_6^{\rm pole}$ are the non-regular ones 
in the soft limit, and to 
order $k_5, k_6$, they read:
\ea{
A_{6, \rm Laurent} &= \sum_{m=1}^3 \frac{A_4(m,5,6,-[m56])A_4({\rm complement})}{s_{m56}}
=
-\frac{\lambda^4}{v^8}\sum_{m=1}^3 \frac{ (k_m k_5)(k_m k_6) }{ k_m (k_5 + k_6) } + \Ord(k_5^2, k_6^2)
}
where by the `complement' we mean the other three momenta of the six-point amplitude on the external legs of $A_4$ and $[m56]$ on the internal leg. These are the lowest order terms in the soft expansion of $A_6^{\rm pole}$, and correspond to the physical case where two soft dilatons are emitted simultaneously from one hard external leg. As such they are trivial.

Focusing on the nontrivial soft part of the six-point amplitude coming from $A_6^{\partial^4}$
it is straightforward to check that
\eas{
\lim_{k_5, k_6 \to 0} &A_{6, \rm Taylor} (1,2,3,\bar{4}, 5,6) =
\lambda^2 \lim_{k_5, k_6 \to 0} A_6^{\partial^4} (1,2,3,\bar{4}, 5,6) = \frac{20}{v^2} A_4 (1,2,3,\bar{4} )
 \\
&=
\frac{1}{v^2} \left ( -1- \sum_{i=1}^3 k_i \cdot \partial_{k_i} \right ) 
\left (- \sum_{i=1}^3 k_i \cdot \partial_{k_i} \right ) A_4 (1,2,3,\bar{4} )
}
 where the last line readily follows from \Eq{leading54}.
 It is likewise easy to check the second double-soft identity.
 \eas{
 \lim_{k_5, k_6 \to 0} \partial_{5,6}^\mu A_{6, \rm Taylor} (1,2,3,\bar{4}, 5,6) &=
\lambda^2 \lim_{k_5, k_6 \to 0} \partial_{5,6}^\mu  A_6^{\partial^4} (1,2,3,\bar{4}, 5,6)
 \\
& = 
-\frac{10\lambda^2}{v^6}
\left ( s_{12} \, k_3^\mu+s_{13}\, k_2^\mu + s_{23} \,k_1^\mu
 \right )
 \\
& = 
\frac{1}{v^2}  \sum_{i=1}^3  \hat{K}_{k_i}^\mu
\left ( -1- \sum_{i=1}^3 k_i \cdot \partial_{k_i} \right ) 
A_4 (1,2,3,\bar{4} )
 }
 where the last line follows immediately from \Eq{leading54} and \eqref{subleading54}.

\section{Conclusions}
\label{conclusions}
In this paper we have studied the Ward identities of spontaneously broken 
scale and special conformal invariance, and from them derived the consequences 
for scattering amplitudes describing the interaction between the dilaton 
(the Nambu-Goldstone boson of the spontaneously broken conformal symmetry)
  and other spinless particles.

We have shown that the Ward identities give rise to soft theorems for the
 dilaton, 
which fix the behavior of scattering amplitudes involving soft dilatons, when
scattering on other spinless states. 
The results are straightforward to generalized to scattering on spin-carrying 
states, namely one should simply include the spin-projection part in the analysis 
of special conformal transformations and amputate correlation functions 
accordingly.

Our main new result is the derivation of a double-soft theorem for the dilaton, 
which extends the single soft theorem found in Ref.~\cite{DiVecchia:2015jaq} to the case of double-soft scattering of dilatons. It turns out that the amplitudes factorize in a soft and a hard part through linear order in the soft dilaton momenta, be there one or two soft dilatons involved.
The soft part is given by operators related to the generators of the dilatation and special conformal transformation acting on the hard part, which is just the amplitude involving only the hard states.
 The new double-soft theorem turns out to be equivalent to performing two single-soft limits one after the other, and  we like to point out that this is different from the case of double-soft scattering of pions. This observation allows us to  propose that multi-soft scattering of dilatons should behave in the same way.

The dilaton soft theorems, being consequences of symmetries, are
 independent of a specific microscopic description and as such are universal. 
This means that any (quantum) theory of spontaneously broken conformal 
symmetry must obey the soft theorems put forward in this work. Consequently, 
this puts constraints on any effective description, for instance on the possible 
interactions and coupling in a low-energy effective action of spontaneously 
broken conformal invariance.
We have specifically demonstrated this by checking explicitly the single- and 
double-soft theorems relating 4-, 5-, and 6-point amplitudes in two models; 
one that is valid semiclassically in any number of dimensions, and another that 
is fully valid in the quantum theory but only in four dimensions; namely the 
Coulomb branch in $\mathcal{N}=4$ supersymmetric Yang-Mills theory, 
which we studied in the strongly coupled regime.
Both theories are frequently studied in the literature, and our detailed checks 
may serve as new relations among amplitudes of the theories that were not 
noticed before.

\vspace{-5mm}
\subsection*{Acknowledgments} \vspace{-3mm}
{We thank Massimo Bianchi, Marialuisa Frau, Andrea Guerrieri,  
Yu-tin Huang, Yegor Korovin,  Alberto Lerda, Rodolfo Russo
 and Congkao Wen for useful discussions.
}
\appendix

\section{
Single-soft limit of $A_6$ of section~\ref{N=4}
}
\label{56dilaton}

In this appendix we show that \Eq{cancellation} is fulfilled.
Let us summarize the expressions for the amplitudes in Sec.~\ref{N=4}:
\ea{
A_4 (1,2,3,4) &= \frac{\lambda^2}{4 v^4} [ s_{12} s_{34} + s_{13} s_{24}+s_{14} s_{23} ] 
 \\[2mm]
 \begin{split}
 A_5(1,2,3,4,5) &= - \frac{4}{v} \Big [ A_4(1,2,3,4)
+A_4(1,2,3,5) + A_4(1,2,4,5)
\\
& \qquad + A_4(1,3,4,5)+A_4(2,3,4,5) \Big]
\end{split}
\\[2mm]
A_6 &= \lambda^2 A_6^{\partial^4} +\lambda^4 
\left (A_6^{\partial^6} + A_6^{\rm pole} \right )
\label{A6vv}}
with
\ea{
\lambda^2 A_6^{\partial^4}&=
\frac{20}{v^2} \sum_{\rm cycl. perm}^{1,\ldots 6} \left [
A_4(1,2,3,4)
+
A_4(1,2,3,5) +
\frac{1}{2} A_4(1,2,4,5)  \right ]
\\
\lambda^4 A_6^{\partial^6}&=
- \frac{3 \lambda^4}{8 v^8} \sum_{\rm cycl. perm}^{1,\ldots 6}\Bigg [
\frac{s_{14}s_{25}s_{36}}{6}
+ \frac{s_{12}s_{34}s_{56}}{3} 
+ \frac{s_{14}s_{23}s_{56}}{2}
+\frac{s_{15}s_{24}s_{36}}{2}
+ s_{13}s_{24}s_{56}
\Bigg]
\\
\lambda^4 A_6^{\rm pole}
&= \sum_{\text{ineq. perm.}} \frac{A_4(1, 2, 3, - [123] ) A_4([123], 4,5,6)}{s_{123}}
}
where the last sum over inequivalent permutations are given by the denominator structures:
\ea{
\{ s_{123}, \ s_{124}, \ s_{125}, \ s_{126}, \ s_{134}, \ s_{135}, \ s_{136}, \ s_{145}, \ s_{146}, \ s_{156} \}
}

As explained in the main text, the soft limit $k_6 \to 0$ of $A_6$ reproduces 
the correct soft theorem, since 
$A_6^{\partial^6}$ and $A_6^{\rm pole}$ both vanish in this limit.
At subleading order they do not vanish, but should instead cancel each 
other, since they cannot contribute to the soft theorem due to the coupling 
being $\lambda^4$, while $A_5$ has {only} terms with coupling $\lambda^2$.
This cancellation can only occur if the denominators in $A_6^{\rm pole}$ cancel out at subleading order.
Let us first show this.

To show that the denominators of $A_6^{\rm pole}$ cancel out at subleading order, we first rewrite all denominators explicitly in terms of $k_6$:
\eas{
&\to \{ s_{456}, \ s_{356}, \ s_{346}, \ s_{126}, \ s_{256}, \ s_{246}, \ s_{136}, \ s_{236}, \ s_{146}, \ s_{156} \}
 \\[2mm]
& \stackrel{k_6 \to 0}{\to} 
 \{ s_{45}, \ s_{35}, \ s_{34}, \ s_{12}, \ s_{25}, \ s_{24}, \ s_{13}, \ s_{23}, \ s_{14}, \ s_{15} \}
}
Now consider the numerator corresponding to the first term above:
\eas{
&A_4 (1,2,3,[456])A_4(-[456], 4,5,6)
= \left (\frac{\lambda^2}{4 v^4} \right )^2 
\\
&\times
[ s_{12} (s_{34} + s_{35} +s_{36} ) + s_{13} (s_{24} + s_{25} + s_{26} ) +
s_{23} (s_{14} + s_{15} + s_{16} )  ]
\\
&\times
[ - (s_{45} + s_{46}) s_{56} - (s_{45} + s_{56}) s_{46}- (s_{46} + s_{56}) s_{45}  ]
}
To linear order in $k_6$ this expression reduces to:
\eas{
&A_4 (1,2,3,[456])A_4(-[456], 4,5,6)
= - \left (\frac{\lambda^2}{4 v^4} \right )^2 
\\
&\times
[ s_{12} (s_{34} + s_{35} ) + s_{13} (s_{24} + s_{25}  ) +
s_{23} (s_{14} + s_{15} )  ]
\\
&\times
2 [  s_{46} +  s_{56}  ] s_{45} + \Ord (k_6^2)
}
We observe that $s_{45}$ factorizes and exactly cancels the denominator, which
is also {equal to} $s_{45}$.
We may also observe that the second line is simply:
\ea{
[ s_{12} (s_{34} + s_{35} ) + s_{13} (s_{24} + s_{25}  ) +
s_{23} (s_{14} + s_{15} )  ] ={ \frac{4v^4}{\lambda^2} \left( A_4(1,2,3,4) + 
A_4 (1,2,3,5) \right)}
}
Summarizing, we have shown that:
\ea{
\frac{A_4(1, 2, 3, [456] ) A_4(-[456], 4,5,6)}{s_{456}}
=
- \frac{\lambda^2}{ v^4} k_6 \cdot (k_4 + k_5) \Big [  A_4(1,2,3,4) + A_4 (1,2,3,5) \Big ] + \Ord(k_6^2)
}
By summing over all ten inequivalent permutation terms we find
(for short {we}  denote $A_4(i,j,k,l) = A_{ijkl}$)
\eas{
\lambda^4 A_6^{\rm pole} = 
&- \frac{\lambda^2}{ v^4} k_6 \cdot k_1
\Big [ A_{1234} + A_{1235} + A_{1245} + A_{1345} + 4 A_{2345} \Big ]
\\
&
- \frac{\lambda^2}{ v^4} k_6 \cdot k_2
\Big [ A_{1234} + A_{1235} + A_{1245} + A_{2345} + 4 A_{1345} \Big ]
 \\
&
- \frac{\lambda^2}{ v^4} k_6 \cdot k_3
\Big [ A_{1234} + A_{1235} + A_{1345} + A_{2345} + 4 A_{1245} \Big ]
  \\
&
- \frac{\lambda^2}{ v^4} k_6 \cdot k_4
\Big [ A_{1234} + A_{1245} + A_{1345} + A_{2345} + 4 A_{1235} \Big ]
  \\
&
- \frac{\lambda^2}{ v^4} k_6 \cdot k_5
\Big [ A_{1235} + A_{1245} + A_{1345} + A_{2345} + 4 A_{1234} \Big ] + \Ord (k_6^2)
  \\
= &- \frac{\lambda^2}{ v^4} k_6 \cdot (k_1 + k_2 + k_3 + k_4 + 4 k_5) A_{1234} + \cdots
  \\
= &- \frac{\lambda^2}{ v^4} k_6 \cdot (3 k_5) A_{1234} + \cdots
}
where the $\cdots$ in the last and 
next to last line should be understood as the 5 other terms, which are 
simply the 5 cyclic permutations of the indices $2345$. To get the last 
expression we used momentum conservation 
$k_1 + k_2 + k_3 + k_4 = - k_5 - k_6$ where $k_6$ 
gives rise to a  higher  order term and can be neglected. 
Notice that we are not fixing one momentum by momentum 
conservation, rather we use it to simplify expressions. One may fix a 
momentum in the end after all rewritings.
Explicitly, we have found:
\ea{
\lambda^4 A_6^{\rm pole}=
- 3 \frac{\lambda^2}{ v^4} \Big [
A_{1234} \, k_5 + A_{1235} \, k_4 + A_{1245}\,  k_3 + A_{1345}\,  k_4 + A_{2345} \, k_1 \Big ] \cdot k_6 + \Ord(k_6^2)
\label{linearA6pole}
}

Let us now consider $A_6^{\partial^6}$ which is linear in $k_6$ (in fact, in any 
momenta):
\eas{
\lambda^4 A_6^{\partial^6}=
 \frac{3 \lambda^4}{8 v^8}&\Bigg [
s_{14}s_{25}s_{36}
+ s_{12}s_{34}s_{56}+ s_{23}s_{45}s_{61}
+ s_{14}s_{23}s_{56}+ s_{25}s_{34}s_{16}
 \\
&
+ s_{36}s_{45}s_{12}
+s_{15}s_{24}s_{36}+ s_{26}s_{35}s_{14}+ s_{13}s_{46}s_{25}
+ s_{13}s_{24}s_{56}
  \\
&
+ s_{24}s_{35}s_{61}
+ s_{35}s_{46}s_{12}
+ s_{46}s_{51}s_{23}
+ s_{51}s_{62}s_{34}
+ s_{62}s_{13}s_{45}
\Bigg]
}
It is easy to see that by factorizing $k_6$ in each term and collecting together the $k_i$ terms it multiplies we get:
\ea{
\lambda^4 A_6^{\partial^6}=
  \frac{3 \lambda^2}{ v^4} 
 \Big [
A_{1234} \, k_5 + A_{1235} \, k_4 + A_{1245}\,  k_3 + A_{1345}\,  k_4 + A_{2345} \, k_1 \Big ] \cdot k_6
}
Comparing this expression with that in \Eq{linearA6pole}, 
we observe that they are identical but with opposite sign.
Thus at linear order in  $k_6$ the 
terms proportional to $\lambda^4$ in Eq. (\ref{A6vv}) do not contribute,
which as explained is an expected consequence of the 
soft theorem at subleading order in the soft momentum $k_6$.
This reversibly illustrates the strong constraints that soft theorems put on effective field theories.


\begin{thebibliography}{99}
\small


%\cite{Nielsen:1975hm}
\bibitem{Nielsen:1975hm} 
  H.~B.~Nielsen and S.~Chadha,
  %``On How to Count Goldstone Bosons,''
  Nucl.\ Phys.\ B {\bf 105}, 445 (1976).
  doi:10.1016/0550-3213(76)90025-0
  %%CITATION = doi:10.1016/0550-3213(76)90025-0;%%
  %128 citations counted in INSPIRE as of 28 Apr 2017


%\cite{Higashijima:1994zg}
\bibitem{Higashijima:1994zg} 
  K.~Higashijima,
  %``Nambu-Goldstone theorem for conformal symmetry,''
  In *Toyonaka 1994, Group theoretical methods in physics* 223-228


%\cite{Low:2001bw}
\bibitem{Low:2001bw} 
  I.~Low and A.~V.~Manohar,
  %``Spontaneously broken space-time symmetries and Goldstone's theorem,''
  Phys.\ Rev.\ Lett.\  {\bf 88}, 101602 (2002)
  doi:10.1103/PhysRevLett.88.101602
  [hep-th/0110285].
  %%CITATION = doi:10.1103/PhysRevLett.88.101602;%%
  %118 citations counted in INSPIRE as of 28 Apr 2017


%\cite{Adler:1964um}
\bibitem{Adler:1964um} 
  S.~L.~Adler,
  %``Consistency conditions on the strong interactions implied by a partially conserved axial vector current,''
  Phys.\ Rev.\  {\bf 137}, B1022 (1965).
  doi:10.1103/PhysRev.137.B1022
  %%CITATION = doi:10.1103/PhysRev.137.B1022;%%
  %411 citations counted in INSPIRE as of 28 Apr 2017


%\cite{Adler:1965ga}
\bibitem{Adler:1965ga} 
  S.~L.~Adler,
  %``Consistency conditions on the strong interactions implied by a partially conserved axial-vector current. II,''
  Phys.\ Rev.\  {\bf 139}, B1638 (1965).
  doi:10.1103/PhysRev.139.B1638
  %%CITATION = doi:10.1103/PhysRev.139.B1638;%%
  %248 citations counted in INSPIRE as of 28 Apr 2017


%\cite{Weinberg:1966gjf}
\bibitem{Weinberg:1966gjf} 
  S.~Weinberg,
  %``Current-Commutator Theory of Multiple Pion Production,''
  Phys.\ Rev.\ Lett.\  {\bf 16}, no. 19, 879 (1966).
  doi:10.1103/PhysRevLett.16.879
  %%CITATION = doi:10.1103/PhysRevLett.16.879;%%
  %94 citations counted in INSPIRE as of 28 Apr 2017


%\cite{Mack:1968zz}
\bibitem{Mack:1968zz} 
  G.~Mack,
  %``Partially conserved dilatation current,''
  Nucl.\ Phys.\ B {\bf 5}, 499 (1968).
  doi:10.1016/0550-3213(68)90232-0
  %%CITATION = doi:10.1016/0550-3213(68)90232-0;%%
  %93 citations counted in INSPIRE as of 28 Apr 2017


%\cite{Gross:1970tb}
\bibitem{Gross:1970tb} 
  D.~J.~Gross and J.~Wess,
  %``Scale invariance, conformal invariance, and the high-energy behavior of scattering amplitudes,''
  Phys.\ Rev.\ D {\bf 2}, 753 (1970).
  doi:10.1103/PhysRevD.2.753
  %%CITATION = doi:10.1103/PhysRevD.2.753;%%
  %84 citations counted in INSPIRE as of 28 Apr 2017


%\cite{Gursey:1956zzb}
\bibitem{Gursey:1956zzb} 
  F.~Gursey,
  %``On a conform-invariant spinor wave equation,''
  Nuovo Cim.\  {\bf 3}, 988 (1956).
  doi:10.1007/BF02823498
  %%CITATION = doi:10.1007/BF02823498;%%
  %50 citations counted in INSPIRE as of 28 Apr 2017

 %\cite{Wess60}
\bibitem{Wess60} 
  J.~Wess,
  %``The Conformal Invariance in Quantum Field Theory''
  Nuovo Cim.\  {\bf 18}, 1086 (1960).
doi:10.1007/BF02733168

%\cite{Kastrup:1962zza}
\bibitem{Kastrup:1962zza} 
  H.~A.~Kastrup,
  %``On the physical interpretation and representation-theoretic analysis of the conformal transformations of space and time,''
  Annalen Phys.\  {\bf 464}, 388 (1962)
  [Annalen Phys.\  {\bf 9}, 388 (1962)].
  doi:10.1002/andp.19624640706
  %%CITATION = doi:10.1002/andp.19624640706;%%
  %57 citations counted in INSPIRE as of 28 Apr 2017


%\cite{Fulton:1962bu}
\bibitem{Fulton:1962bu} 
  T.~Fulton, F.~Rohrlich and L.~Witten,
  %``Conformal invariance in physics,''
  Rev.\ Mod.\ Phys.\  {\bf 34}, 442 (1962).
  doi:10.1103/RevModPhys.34.442
  %%CITATION = doi:10.1103/RevModPhys.34.442;%%
  %154 citations counted in INSPIRE as of 28 Apr 2017


%\cite{Callan:1970yg}
\bibitem{Callan:1970yg} 
  C.~G.~Callan, Jr.,
  %``Broken scale invariance in scalar field theory,''
  Phys.\ Rev.\ D {\bf 2}, 1541 (1970).
  doi:10.1103/PhysRevD.2.1541
  %%CITATION = doi:10.1103/PhysRevD.2.1541;%%
  %980 citations counted in INSPIRE as of 28 Apr 2017


%\cite{Coleman:1970je}
\bibitem{Coleman:1970je} 
  S.~R.~Coleman and R.~Jackiw,
  %``Why dilatation generators do not generate dilatations?,''
  Annals Phys.\  {\bf 67}, 552 (1971).
  doi:10.1016/0003-4916(71)90153-9
  %%CITATION = doi:10.1016/0003-4916(71)90153-9;%%
  %311 citations counted in INSPIRE as of 28 Apr 2017


%\cite{Boels:2015pta}
\bibitem{Boels:2015pta} 
  R.~H.~Boels and W.~Wormsbecher,
  %``Spontaneously broken conformal invariance in observables,''
  arXiv:1507.08162 [hep-th].
  %%CITATION = ARXIV:1507.08162;%%
  %11 citations counted in INSPIRE as of 28 Apr 2017


%\cite{Huang:2015sla}
\bibitem{Huang:2015sla} 
  Y.~t.~Huang and C.~Wen,
  %``Soft theorems from anomalous symmetries,''
  JHEP {\bf 1512}, 143 (2015)
  doi:10.1007/JHEP12(2015)143
  [arXiv:1509.07840 [hep-th]].
  %%CITATION = doi:10.1007/JHEP12(2015)143;%%
  %11 citations counted in INSPIRE as of 28 Apr 2017


%\cite{DiVecchia:2015jaq}
\bibitem{DiVecchia:2015jaq} 
  P.~Di Vecchia, R.~Marotta, M.~Mojaza and J.~Nohle,
  %``New soft theorems for the gravity dilaton and the Nambu-Goldstone dilaton at subsubleading order,''
  Phys.\ Rev.\ D {\bf 93}, no. 8, 085015 (2016)
  doi:10.1103/PhysRevD.93.085015
  [arXiv:1512.03316 [hep-th]].
  %%CITATION = doi:10.1103/PhysRevD.93.085015;%%
  %12 citations counted in INSPIRE as of 28 Apr 2017


%\cite{Low:1958sn}
\bibitem{Low:1958sn} 
  F.~E.~Low,
  %``Bremsstrahlung of very low-energy quanta in elementary particle collisions,''
  Phys.\ Rev.\  {\bf 110}, 974 (1958).
  doi:10.1103/PhysRev.110.974
  %%CITATION = doi:10.1103/PhysRev.110.974;%%
  %723 citations counted in INSPIRE as of 28 Apr 2017


%\cite{Weinberg:1964ew}
\bibitem{Weinberg:1964ew} 
  S.~Weinberg,
  %``Photons and Gravitons in s Matrix Theory: Derivation of Charge Conservation and Equality of Gravitational and Inertial Mass,''
  Phys.\ Rev.\  {\bf 135}, B1049 (1964).
  doi:10.1103/PhysRev.135.B1049
  %%CITATION = doi:10.1103/PhysRev.135.B1049;%%
  %339 citations counted in INSPIRE as of 28 Apr 2017




%\cite{Antipin:2011aa}
\bibitem{Antipin:2011aa} 
  O.~Antipin, M.~Mojaza and F.~Sannino,
  %``Light Dilaton at Fixed Points and Ultra Light Scale Super Yang Mills,''
  Phys.\ Lett.\ B {\bf 712}, 119 (2012)
  doi:10.1016/j.physletb.2012.04.050
  [arXiv:1107.2932 [hep-ph]].
  %%CITATION = doi:10.1016/j.physletb.2012.04.050;%%
  %42 citations counted in INSPIRE as of 28 Apr 2017

%\cite{Bianchi:2016viy}
\bibitem{Bianchi:2016viy} 
  M.~Bianchi, A.~L.~Guerrieri, Y.~t.~Huang, C.~J.~Lee and C.~Wen,
  %``Exploring soft constraints on effective actions,''
  JHEP {\bf 1610}, 036 (2016)
  doi:10.1007/JHEP10(2016)036
  [arXiv:1605.08697 [hep-th]].
  %%CITATION = doi:10.1007/JHEP10(2016)036;%%
  %7 citations counted in INSPIRE as of 28 Apr 2017



%\cite{Dashen:1969ez}
\bibitem{Dashen:1969ez} 
  R.~F.~Dashen and M.~Weinstein,
  %``Soft pions, chiral symmetry, and phenomenological lagrangians,''
  Phys.\ Rev.\  {\bf 183}, 1261 (1969).
  doi:10.1103/PhysRev.183.1261
  %%CITATION = doi:10.1103/PhysRev.183.1261;%%
  %188 citations counted in INSPIRE as of 28 Apr 2017


%\cite{ArkaniHamed:2008gz}
\bibitem{ArkaniHamed:2008gz} 
  N.~Arkani-Hamed, F.~Cachazo and J.~Kaplan,
  %``What is the Simplest Quantum Field Theory?,''
  JHEP {\bf 1009}, 016 (2010)
  doi:10.1007/JHEP09(2010)016
  [arXiv:0808.1446 [hep-th]].
  %%CITATION = doi:10.1007/JHEP09(2010)016;%%
  %359 citations counted in INSPIRE as of 28 Apr 2017


%\cite{Kampf:2013vha}
\bibitem{Kampf:2013vha} 
  K.~Kampf, J.~Novotny and J.~Trnka,
  %``Tree-level Amplitudes in the Nonlinear Sigma Model,''
  JHEP {\bf 1305}, 032 (2013)
  doi:10.1007/JHEP05(2013)032
  [arXiv:1304.3048 [hep-th]].
  %%CITATION = doi:10.1007/JHEP05(2013)032;%%
  %27 citations counted in INSPIRE as of 28 Apr 2017


%\cite{Low:2015ogb}
\bibitem{Low:2015ogb} 
  I.~Low,
  %``Double Soft Theorems and Shift Symmetry in Nonlinear Sigma Models,''
  Phys.\ Rev.\ D {\bf 93}, no. 4, 045032 (2016)
  doi:10.1103/PhysRevD.93.045032
  [arXiv:1512.01232 [hep-th]].
  %%CITATION = doi:10.1103/PhysRevD.93.045032;%%
  %12 citations counted in INSPIRE as of 28 Apr 2017


%\cite{Du:2015esa}
\bibitem{Du:2015esa} 
  Y.~J.~Du and H.~Luo,
  %``On single and double soft behaviors in NLSM,''
  JHEP {\bf 1508}, 058 (2015)
  doi:10.1007/JHEP08(2015)058
  [arXiv:1505.04411 [hep-th]].
  %%CITATION = doi:10.1007/JHEP08(2015)058;%%
  %10 citations counted in INSPIRE as of 28 Apr 2017


%\cite{Strominger:2013lka}
\bibitem{Strominger:2013lka} 
  A.~Strominger,
  %``Asymptotic Symmetries of Yang-Mills Theory,''
  JHEP {\bf 1407}, 151 (2014)
  doi:10.1007/JHEP07(2014)151
  [arXiv:1308.0589 [hep-th]].
  %%CITATION = doi:10.1007/JHEP07(2014)151;%%
  %86 citations counted in INSPIRE as of 28 Apr 2017


%\cite{Strominger:2013jfa}
\bibitem{Strominger:2013jfa} 
  A.~Strominger,
  %``On BMS Invariance of Gravitational Scattering,''
  JHEP {\bf 1407}, 152 (2014)
  doi:10.1007/JHEP07(2014)152
  [arXiv:1312.2229 [hep-th]].
  %%CITATION = doi:10.1007/JHEP07(2014)152;%%
  %184 citations counted in INSPIRE as of 28 Apr 2017


%\cite{Campiglia:2017dpg}
\bibitem{Campiglia:2017dpg} 
  M.~Campiglia, L.~Coito and S.~Mizera,
  %``Can scalars have asymptotic symmetries?,''
  arXiv:1703.07885 [hep-th].
  %%CITATION = ARXIV:1703.07885;%%
  %3 citations counted in INSPIRE as of 28 Apr 2017


%\cite{Luo:2015tat}
\bibitem{Luo:2015tat} 
  H.~Luo and C.~Wen,
  %``Recursion relations from soft theorems,''
  JHEP {\bf 1603}, 088 (2016)
  doi:10.1007/JHEP03(2016)088
  [arXiv:1512.06801 [hep-th]].
  %%CITATION = doi:10.1007/JHEP03(2016)088;%%
  %14 citations counted in INSPIRE as of 28 Apr 2017


%\cite{Strominger:2017zoo}
\bibitem{Strominger:2017zoo} 
  A.~Strominger,
  %``Lectures on the Infrared Structure of Gravity and Gauge Theory,''
  arXiv:1703.05448 [hep-th].
  %%CITATION = ARXIV:1703.05448;%%
  %9 citations counted in INSPIRE as of 28 Apr 2017
  

%\cite{Ademollo:1975pf}
\bibitem{Ademollo:1975pf} 
 M. Ademollo, A. D'Adda, R. D'Auria, F. Gliozzi, E. Napolitano,  S. Sciuto and 
P. Di Vecchia, Nucl. Phys. {\bf B94}, (1975) 221 doi:10.1016/0550-3213(75)90491-5
%\mbox{\href{http://dx.doi.org/10.1016/0550-3213(75)90491-5}{{\em 
%``Soft Dilatons and Scale Renormalization in Dual Theories,''
%Nucl.\ Phys.\ } {\bf B94}, (1975) 221}};
  %%CITATION = doi:10.1016/0550-3213(75)90491-5;%%
  %
 
  %\cite{Shapiro:1975cz}
 \bibitem{Shapiro:1975cz}
%``On the renormalization of Dual Models'',
J. Shapiro,  Phys. Rev {\bf D11} (1975) 2937 doi:.1103/PhysRevD.11.2937.
%\href{http://dx.doi.org/ 10.1103/PhysRevD.11.2937}{{\em Phys.
%Rev.} {\bf D11} (1975) 2937}.
  %%CITATION = doi:10.1103/PhysRevD.11.2937;%%
  
  %\cite{DiVecchia:2015oba,DiVecchia:2016amo}
 \bibitem{DiVecchia:2015oba} 
   P.~Di Vecchia, R.~Marotta and M.~Mojaza, JHEP {\bf 1505}, 137 (2015) doi:10.1007/JHEP05(2015)137 [arXiv:1502.05258 [hep-th]].
   %``Soft theorem for the graviton, dilaton and the Kalb-Ramond field in the bosonic string,''
% \href{http://dx.doi.org/ 10.1007/JHEP05(2015)137}{{\em  JHEP} {\bf 1505}, 137 (2015)}
% \href{http://arxiv.org/abs/1502.05258}{{\ttfamily  arXiv:1502.05258}}.
   %%CITATION = ARXIV:1502.05258;%%
   %8 citations counted in INSPIRE as of 13 Nov 2015
 
%\cite{DiVecchia:2016amo}
\bibitem{DiVecchia:2016amo} 
  P.~Di Vecchia, R.~Marotta and M.~Mojaza,
  JHEP {\bf 1606}, 054 (2016)  doi:10.1007/10.1007/JHEP06(2016)054, [arXiv:1604.03355 [hep-th]].
  %``Subsubleading soft theorems of gravitons and dilatons in the bosonic string,''
%  \href{http://dx.doi.org/ 10.1007/10.1007/JHEP06(2016)054}{{\em JHEP} {\bf 1606}, 054 (2016)}, \href{http://arxiv.org/abs/1604.03355}{{\ttfamily  arXiv:1604.03355}}.
  %%CITATION = doi:10.1007/JHEP06(2016)054;%%
   %
   
   %\cite{DiVecchia:2016szw}
\bibitem{DiVecchia:2016szw} 
  P.~Di Vecchia, R.~Marotta and M.~Mojaza,
  %``Soft behavior of a closed massless state in superstring and universality in the soft behavior of the dilaton,''
  JHEP {\bf 1612}, 020 (2016)
  doi:10.1007/JHEP12(2016)020
  [arXiv:1610.03481 [hep-th]].
  %%CITATION = doi:10.1007/JHEP12(2016)020;%%
  %5 citations counted in INSPIRE as of 28 Apr 2017



%\cite{Mack:1969rr}
\bibitem{Mack:1969rr} 
  G.~Mack and A.~Salam,
  %``Finite component field representations of the conformal group,''
  Annals Phys.\  {\bf 53}, 174 (1969).
  doi:10.1016/0003-4916(69)90278-4
  %%CITATION = doi:10.1016/0003-4916(69)90278-4;%%
  %493 citations counted in INSPIRE as of 28 Apr 2017


%\cite{Callan:1970ze}
\bibitem{Callan:1970ze} 
  C.~G.~Callan, Jr., S.~R.~Coleman and R.~Jackiw,
  %``A New improved energy - momentum tensor,''
  Annals Phys.\  {\bf 59}, 42 (1970).
  doi:10.1016/0003-4916(70)90394-5
  %%CITATION = doi:10.1016/0003-4916(70)90394-5;%%
  %827 citations counted in INSPIRE as of 28 Apr 2017


%\cite{DiFrancesco:1997nk}
\bibitem{DiFrancesco:1997nk} 
  P.~Di Francesco, P.~Mathieu and D.~Senechal,
  %``Conformal Field Theory,''
  doi:10.1007/978-1-4612-2256-9
  %%CITATION = doi:10.1007/978-1-4612-2256-9;%%
  %78 citations counted in INSPIRE as of 28 Apr 2017


%\cite{Goldberger:2008zz}
\bibitem{Goldberger:2008zz} 
  W.~D.~Goldberger, B.~Grinstein and W.~Skiba,
  %``Distinguishing the Higgs boson from the dilaton at the Large Hadron Collider,''
  Phys.\ Rev.\ Lett.\  {\bf 100}, 111802 (2008)
  doi:10.1103/PhysRevLett.100.111802
  [arXiv:0708.1463 [hep-ph]].
  %%CITATION = doi:10.1103/PhysRevLett.100.111802;%%
  %247 citations counted in INSPIRE as of 28 Apr 2017


%\cite{Schwimmer:2010za}
\bibitem{Schwimmer:2010za} 
  A.~Schwimmer and S.~Theisen,
  %``Spontaneous Breaking of Conformal Invariance and Trace Anomaly Matching,''
  Nucl.\ Phys.\ B {\bf 847}, 590 (2011)
  doi:10.1016/j.nuclphysb.2011.02.003
  [arXiv:1011.0696 [hep-th]].
  %%CITATION = doi:10.1016/j.nuclphysb.2011.02.003;%%
  %52 citations counted in INSPIRE as of 28 Apr 2017


%\cite{Komargodski:2011vj}
\bibitem{Komargodski:2011vj} 
  Z.~Komargodski and A.~Schwimmer,
  %``On Renormalization Group Flows in Four Dimensions,''
  JHEP {\bf 1112}, 099 (2011)
  doi:10.1007/JHEP12(2011)099
  [arXiv:1107.3987 [hep-th]].
  %%CITATION = doi:10.1007/JHEP12(2011)099;%%
  %316 citations counted in INSPIRE as of 28 Apr 2017


%\cite{Elvang:2012st}
\bibitem{Elvang:2012st} 
  H.~Elvang, D.~Z.~Freedman, L.~Y.~Hung, M.~Kiermaier, R.~C.~Myers and S.~Theisen,
  %``On renormalization group flows and the a-theorem in 6d,''
  JHEP {\bf 1210}, 011 (2012)
  doi:10.1007/JHEP10(2012)011
  [arXiv:1205.3994 [hep-th]].
  %%CITATION = doi:10.1007/JHEP10(2012)011;%%
  %73 citations counted in INSPIRE as of 28 Apr 2017


%\cite{Elvang:2012yc}
\bibitem{Elvang:2012yc} 
  H.~Elvang and T.~M.~Olson,
  %``RG flows in d dimensions, the dilaton effective action, and the a-theorem,''
  JHEP {\bf 1303}, 034 (2013)
  doi:10.1007/JHEP03(2013)034
  [arXiv:1209.3424 [hep-th]].
  %%CITATION = doi:10.1007/JHEP03(2013)034;%%
  %25 citations counted in INSPIRE as of 28 Apr 2017


%\cite{Shaposhnikov:2009nk}
\bibitem{Shaposhnikov:2009nk} 
  M.~E.~Shaposhnikov and F.~V.~Tkachov,
  %``Quantum scale-invariant models as effective field theories,''
  arXiv:0905.4857 [hep-th].
  %%CITATION = ARXIV:0905.4857;%%
  %24 citations counted in INSPIRE as of 28 Apr 2017


%\cite{Armillis:2013wya}
\bibitem{Armillis:2013wya} 
  R.~Armillis, A.~Monin and M.~Shaposhnikov,
  %``Spontaneously Broken Conformal Symmetry: Dealing with the Trace Anomaly,''
  JHEP {\bf 1310}, 030 (2013)
  doi:10.1007/JHEP10(2013)030
  [arXiv:1302.5619 [hep-th]].
  %%CITATION = doi:10.1007/JHEP10(2013)030;%%
  %23 citations counted in INSPIRE as of 28 Apr 2017


%\cite{Gretsch:2013ooa}
\bibitem{Gretsch:2013ooa} 
  F.~Gretsch and A.~Monin,
  %``Perturbative conformal symmetry and dilaton,''
  Phys.\ Rev.\ D {\bf 92}, no. 4, 045036 (2015)
  doi:10.1103/PhysRevD.92.045036
  [arXiv:1308.3863 [hep-th]].
  %%CITATION = doi:10.1103/PhysRevD.92.045036;%%
  %26 citations counted in INSPIRE as of 28 Apr 2017


%\cite{Englert:1976ep}
\bibitem{Englert:1976ep} 
  F.~Englert, C.~Truffin and R.~Gastmans,
  %``Conformal Invariance in Quantum Gravity,''
  Nucl.\ Phys.\ B {\bf 117}, 407 (1976).
  doi:10.1016/0550-3213(76)90406-5
  %%CITATION = doi:10.1016/0550-3213(76)90406-5;%%
  %138 citations counted in INSPIRE as of 28 Apr 2017


%\cite{Treiman:1986ep}
\bibitem{Treiman:1986ep} 
  S.~B.~Treiman, E.~Witten, R.~Jackiw and B.~Zumino,
  %``Current Algebra And Anomalies,''
  Singapore: World Scientific (1985)
  %6 citations counted in INSPIRE as of 28 Apr 2017


%\cite{Gross:1970ee}
\bibitem{Gross:1970ee} 
  D.~J.~Gross and R.~Jackiw,
  %``Construction of covariant and gauge invariant t* products,''
  Nucl.\ Phys.\ B {\bf 14}, 269 (1969).
  doi:10.1016/0550-3213(69)90207-7
  %%CITATION = doi:10.1016/0550-3213(69)90207-7;%%
  %68 citations counted in INSPIRE as of 28 Apr 2017 
 
 \end{thebibliography}
\end{document}